\documentclass[fleqn,10pt]{wlscirep}
\usepackage[utf8]{inputenc}
\usepackage[T1]{fontenc}

\title{Mapping the spectrum of 3D communities in human chromosome conformation capture data}

\author[1,$\ast$]{Sang Hoon Lee}
\author[2]{Yeonghoon Kim}
\author[3]{Sungmin Lee}
\author[4]{Xavier Durang}
\author[5]{Per Stenberg}
\author[2,$\dagger$]{Jae-Hyung Jeon}
\author[6,$\ddagger$]{Ludvig Lizana}
\affil[1]{Department of Liberal Arts, Gyeongnam National University of Science and Technology, Jinju 52725, Korea}
\affil[2]{Department of Physics, Pohang University of Science and Technology, Pohang 37673, Korea}
\affil[3]{Department of Physics, Korea University, Seoul 02841, Korea}
\affil[4]{Department of Physics, University of Seoul, Seoul 02504, Korea}
\affil[5]{Department of Ecology and Environmental Science (EMG), Ume{\aa} University, Ume{\aa} 90187, Sweden}
\affil[6]{Integrated Science Lab, Department of Physics, Ume{\aa} University, Ume{\aa} 90187, Sweden}

\affil[$\ast$]{lshlj82@gntech.ac.kr}
\affil[$\dagger$]{jeonjh@postech.ac.kr}
\affil[$\ddagger$]{ludvig.lizana@umu.se}

\begin{abstract}
Several experiments show that the three dimensional (3D) organization of chromosomes affects genetic processes such as transcription and gene regulation. To better understand this connection, researchers developed the Hi-C method that is able to detect the pairwise physical contacts of all chromosomal loci. The Hi-C data show that chromosomes are composed of 3D compartments that range over a variety of scales. However, it is challenging to systematically detect these cross-scale structures. Most studies have therefore designed methods for specific scales to study foremost topologically associated domains (TADs) and A/B compartments. To go beyond this limitation, we tailor a network community detection method that finds communities in compact fractal globule polymer systems. Our method allows us to continuously scan through all scales with a single resolution parameter. We found: ($i$) polymer segments belonging to the same 3D community do not have to be in consecutive order along the polymer chain. In other words, several TADs may belong to the same 3D community. ($ii$) CTCF proteins---a loop-stabilizing protein that is ascribed a big role in TAD formation---are well correlated with community borders only at one level of organization. ($iii$) TADs and A/B compartments are traditionally treated as two weakly related 3D structures and detected with different algorithms. With our method, we detect both by simply adjusting the resolution parameter. We therefore argue that they represent two specific levels of a continuous spectrum 3D communities, rather than seeing them as different structural entities.
\end{abstract}
\begin{document}

\flushbottom
\maketitle
%
%
\thispagestyle{empty}


\section*{Introduction}

Experiments that detect the physical contacts between DNA loci in the nucleus, such as Hi-C~\cite{Lieberman-Aiden2009,Rao2014}, show that DNA in the nucleus is not a randomly folded polymer. Rather, across cell types and organisms, Hi-C experiments reveal that chromosomes are built up by a network of three dimensional (3D) compartments. 

At the mega($10^6$)-base-pair scale, two types of coexisting structures stand out. In one, all chromosome loci belong to one of the two so-called A/B compartments, where the chromatin in one compartment is generally more open, accessible, and actively transcribed than the other. In the second type, linear subsections of the genome assemble into topological domains~\cite{Dixon2012,Nora2012}, often referred to as topologically associated domains (TADs)~\cite{Dixon2012,Nora2012,Rao2014}. Plotting Hi-C data as a heat map, TADs show up as local regions with sharp borders with more internal than external contacts. The positions of these borders are correlated with several genetic processes, such as transcription, localization of some epigenetic marks, and DNA-binding profiles of several proteins---most notably CTCF and cohesin~\cite{Dixon2012,Nora2012}.  

The methods that detect TADs are not the same as those that find A/B compartments. Therefore, TADs and A/B compartments are treated as different 3D structures that are only weakly related to each other. Just as for TADs, there are several algorithms tailored for detecting A/B compartments~\cite{Lieberman-Aiden2009,dixon2015chromatin}, each with their strengths and weaknesses. 

To algorithmically detect TAD-like structures, there exists by now a menagerie of network~\cite{Boulos2013,Cabreros2015,YXRWang2017,Sarnataro2017,Belyaeva2017} and clustering approaches~\cite{yu2017identifying,weinreb2015identification,haddad2017ic,KKYan2017,Norton2018,Ball2011,Gopalan2013}. Arguably these methods yield overlapping results, but it is unclear by how much. In particular, some methods cannot deal with TAD-within-TAD hierarchies that become apparent when zooming in TADs in highly resolved Hi-C maps. This means that there is not a universal definition for what a TAD really is. 

Some network approaches are based on community detection methods that are related to what we use here. In Cabreros \emph{et al.}~\cite{Cabreros2015}, the authors suggest a method based on the stochastic block model~\cite{Ball2011,Gopalan2013}, which is another side of the network community detection field  compared to the modularity maximization method that we use here (but as shown recently~\cite{Newman2016}, they are connected). However, there are some limitations in their approach. For example, they binarize the Hi-C data (`no connection' or `connection' with a rather arbitrarily chosen thresholds) thereby discarding contact frequency variations in the Hi-C data. Furthermore, there is no comparative study connecting their communities to biological factors or any mechanistic models. Wang \emph{et al}.~\cite{YXRWang2017} takes a step in this direction, but the method to detect the TADs itself relies on biological factor data and nontrivial threshold criteria. 

To overcome some of these problems, we start by acknowledging that the chromosomes have a richer 3D organization than simply TADs and A/B compartments. These are just two examples. To capture this, we develop network-based method that allows us to scan through 3D structures on all scales with a resolution parameter. In particular, our approach is based on the GenLouvain method, originally designed for network community detection. For a specific value of the resolution parameter, the  method finds the optimal community structure with respect to a null model of the network that has to be specified beforehand. Based on the physics of compact polymer globules, we put forward a null model that is consistent with the average contact probabilities in real Hi-C data~\cite{Lieberman-Aiden2009}. This goes beyond previous Louvain-like studies~\cite{KKYan2017,Norton2018} that treat the Hi-C data as a network with random connections. 

Furthermore, most studies, such as Yan \emph{et al}. and Norton \emph{et al}.~\cite{KKYan2017,Norton2018}, treat TADs as linear contiguous sequences of chromatin. This restriction overrides the GenLouvain algorithm's ability to find the (not necessarily contiguous) optimal community structure in the data set~\cite{Porter2009,Fortunato2010}. We remove this restriction in our study. Therefore, to reduce confusion, we will not use the term TAD, but rather the \emph{3D community} for the cluster of nodes that comes out of the GenLouvain algorithm since they are not necessarily linear contiguous sequences.

\section*{Methods}

We use the GenLouvain algorithm~\cite{GenLouvain} to detect communities in the Hi-C maps (this is an extension of the original Louvain method~\cite{Blondel2008}). This algorithm offers the possibility to find communities at several scales with a single resolution parameter $\gamma$. Contrasting other methods with similar features, the spectrum of communities that we detect is not necessarily hierarchical or nested, as in e.g. Wang \emph{et al}., Fraser \emph{et al}., and Bianco \emph{et al}.~\cite{YXRWang2017,Fraser2015,Bianco2017}. Instead, two different values of $\gamma$ give two different collections of communities, and these do not necessarily have anything to do with each other.

To find the communities, the GenLouvain method tries putting the nodes into different communities to  maximize the so-called modularity function $Q$. This function quantifies by how much more dense the connections are within the communities compared to what they would be in a particular type of network, say a random network. GenLouvain's modularity function is
\begin{equation}
Q = \frac{1}{2m} \sum_{i \ne j} \left[ \left( A_{ij} - \gamma P_{ij}^{\rm null} \right) \delta(g_i,g_j) \right ] \,,
\label{eq:modularity}
\end{equation}
where the sum runs over all nodes in the network, in our case genomic loci, and $A_{ij}$ is the number of contacts between all node pairs $i$ and $j$ ($A_{ij}$ is the adjacency matrix in standard network terminology). The resolution parameter $\gamma$ controls the overall scale of communities we would like to detect, where larger values of $\gamma$ correspond to smaller communities. Furthermore, $P_{ij}^{\rm null}$ is the null-model term that is network-type-specific, and the difference $A_{ij} - P_{ij}^{\rm null}$ therefore measures how strongly nodes $i$ and $j$ are connected in the real network, compared to how strongly we expect them to be given $P_{ij}^{\rm null}$. Finally, $\delta(g_i,g_j)$ is the Kronecker delta that is unity only if nodes $i$ and $j$ belong to the same community (otherwise it is zero), and $m$ is a normalization factor so that $Q$ goes between $-1$ and $+1$.

One of the most popular choices for $P_{ij}^{\rm null}$ is the Newman-Girvan (NG) null-model term for a random network,  $P_{ij}^{\mathrm{NG}} = k_i k_j/(2m)$~\cite{Newman2004}. For a unweighted network where $A_{ij}$ is either zero or one, $k_i$ and $k_j$ are the number of links for nodes $i$ and $j$ ($k_i = \sum_j A_{ij}$). Simply put, the NG null-model assumes that the probability that $i$ and $j$ are connected is proportional to the product of their number of links. The same interpretation holds for weighted networks where $k_i$ becomes the sum of weights on the edges connected to $i$ (``strength'' in network terminology).

However, the NG null-model is too rough an approximation to find communities in Hi-C maps, because it does not obey the well-established contact patterns that we know exist in DNA, or in fact any polymer system. For example, DNA is a long polymer where nodes are arranged in a linear sequence and then folded in 3D. This sets limitations for how frequently two pieces of DNA, or nodes, can join in 3D space (this is usually not a restriction in most networks). As was first discovered in Lieberman-Aiden \emph{et al}.~\cite{Lieberman-Aiden2009}, and many papers thereafter~\cite{burton2014species,yu2012spatial}, the contact probability between two nodes $i$ and $j$ decays as a power-law with the linear distance between them, that is $ \propto|i-j|^{-\alpha}$. Based on this, we propose the following null-model:
\begin{equation}
 P_{ij}^{\mathrm{FG}} = \frac{2m k_i k_j |i-j|^{-\alpha}}{\sum_{i' \ne j'} k_{i'} k_{j'} |i' - j'|^{-\alpha}} \,.
\label{eq:Q_FG}
\end{equation}

The value of the decay exponent $\alpha$ is debatable, but we use $\alpha=1$. There are two main reasons. First, at the mega-base-pair scale of human DNA, the Hi-C data suggest that it is close to one~\cite{Lieberman-Aiden2009} ($\alpha$ is also close to one in mice~\cite{yu2012spatial}). Second, in the next section we will study community detection in the fractal globule polymer (hence the superscript FG in $P_{ij}^{\rm FG}$) where $\alpha=1$~\cite{Tamm2015,Mirny2011}. Nonetheless, we point out that our method does note rely on this choice, and $\alpha$ is in principle a free parameter.

\subsection*{3D communities in fractal globules}
\label{sec:fractal_globule}

Before we investigate the human Hi-C data, we wish to better understand the types of 3D structures that our community detection method picks up. To do this, we use computer-generated fractal globule polymers (denoted by the ``crumpled globule'' in the original article~\cite{grosberg1993crumpled}). This is a compact polymer that mimics the large scale structure of human chromosomes, in particular the scaling relations of the end-to-end distance and contact frequency~\cite{Lieberman-Aiden2009}. The advantage of this approach is that we have the explicit 3D coordinates for every part of the polymer (because we made it), which allows us to visualize and analyze the 3D structure of the communities that we detect. For the Hi-C map, we only have pairwise contact frequencies.

\begin{figure}[ht]
\centering
\begin{tabular}{ll}
(\textbf{a}) & (\textbf{b}) \\
\includegraphics[width=0.3\textwidth]{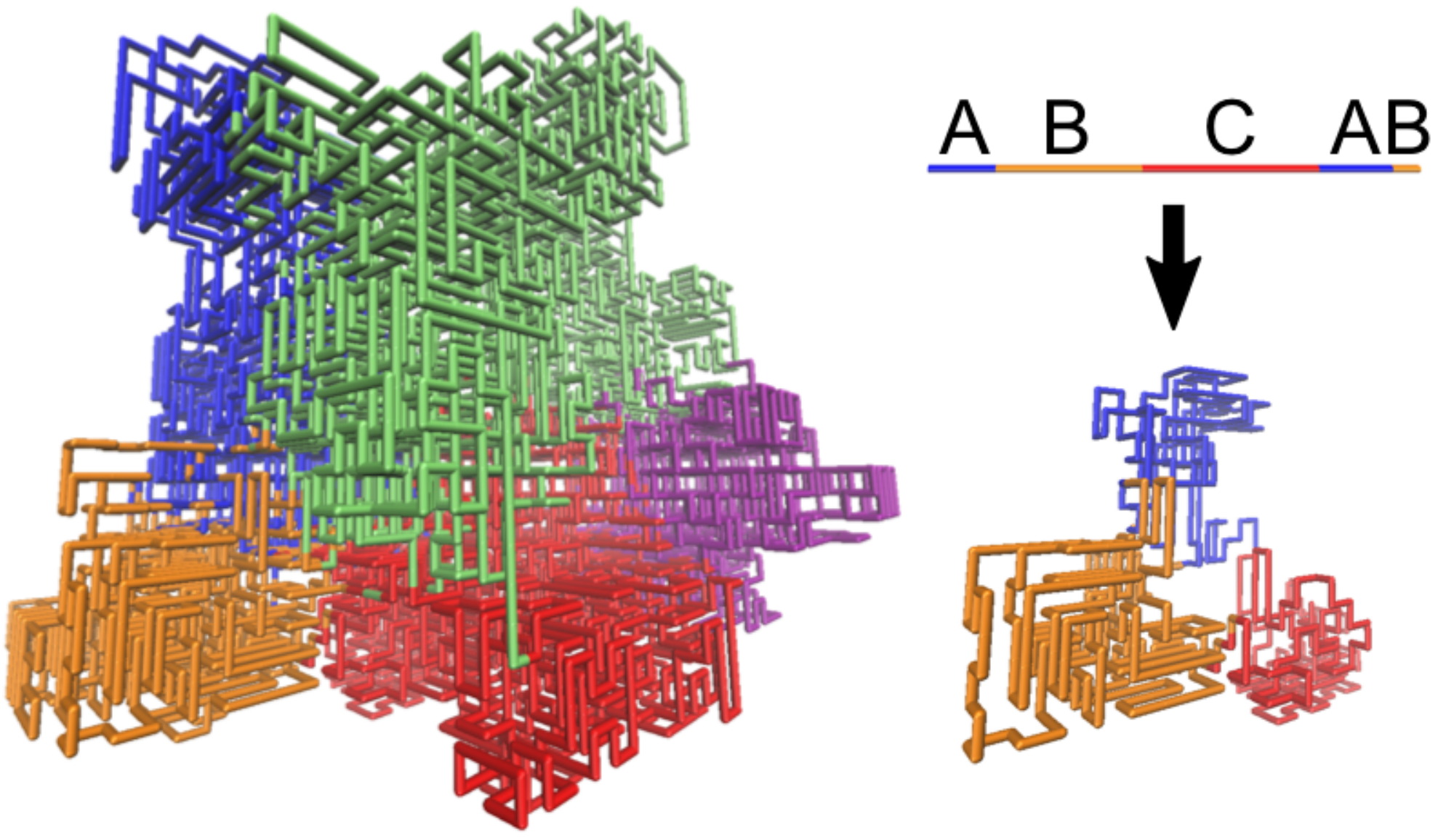} &
\includegraphics[width=0.3\textwidth]{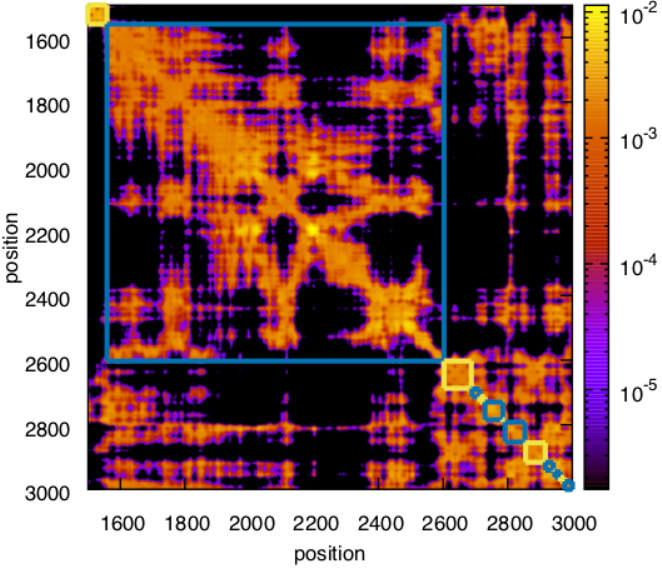} \\
(\textbf{c}) & (\textbf{d}) \\
\includegraphics[width=0.3\textwidth]{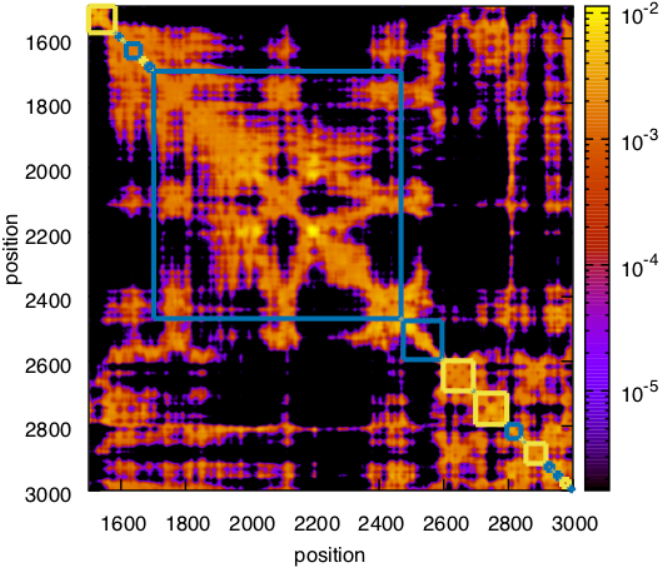} &
\includegraphics[width=0.3\textwidth]{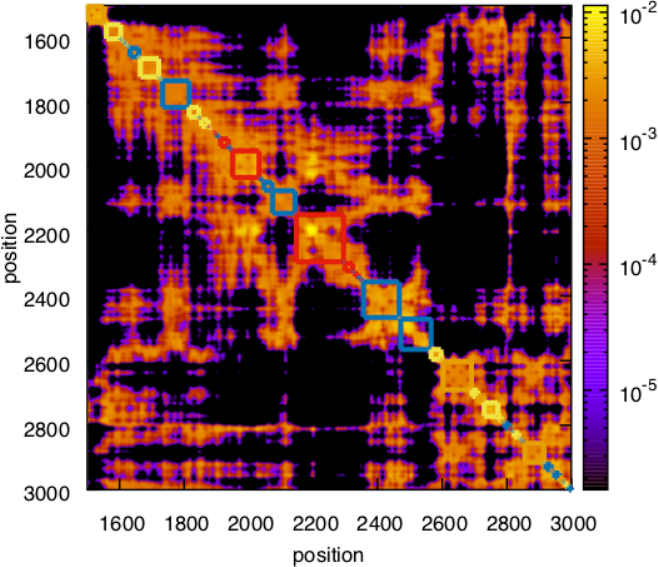} \\
(\textbf{e}) & (\textbf{f}) \\
\includegraphics[width=0.3\textwidth]{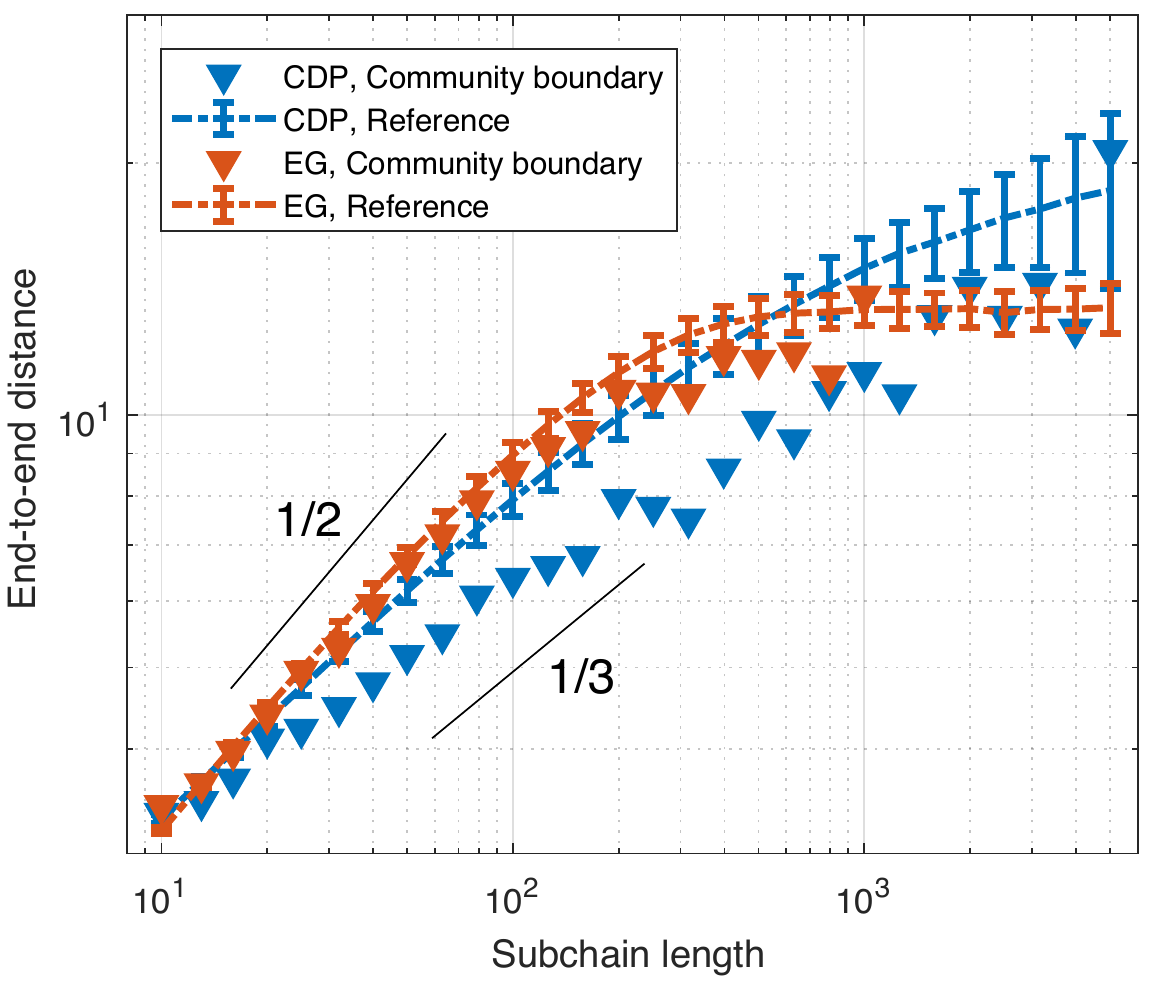} &
\includegraphics[width=0.3\textwidth]{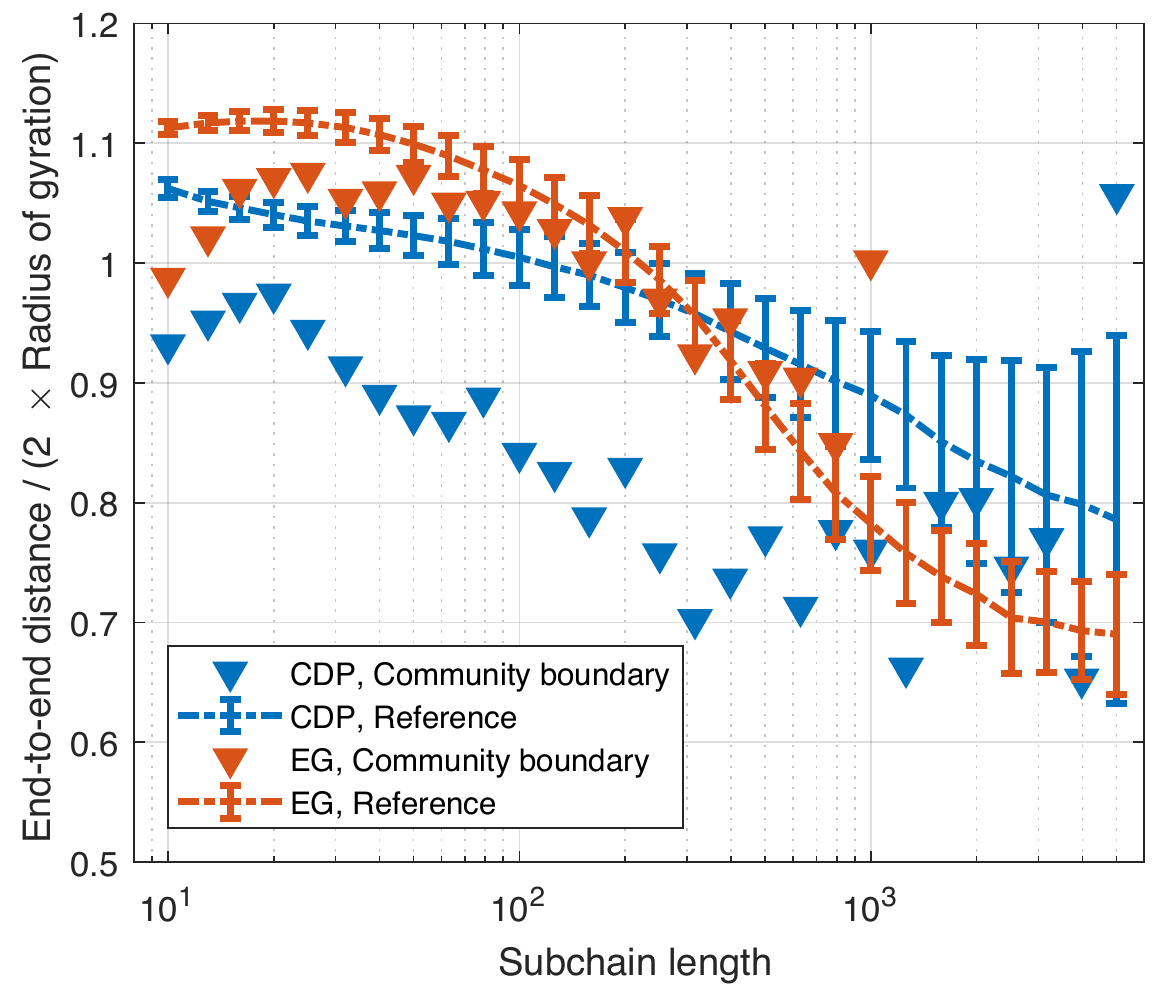} \\
\end{tabular}
\caption{3D communities in simulated fractal globules. (\textbf{a}, left) The 3D representation of fractal globule from the CDP algorithm. The colors highlight communities that we detect using Eq.~\eqref{eq:Q_FG}, $\gamma = 0.6$. (\textbf{a}, right) The communities are not contiguous: the small globule is a subsection of the lower left part of polymer, and the stretched version shows the alternating communities (ABCAB). (\textbf{b})-(\textbf{d}) The contact maps of simulated fractal globules after KR-normalization, with various resolution parameters: (\textbf{b}) $\gamma = 0.4$, (\textbf{c}) $\gamma = 0.6$, and (\textbf{d}) $\gamma = 0.8$. To show non-contiguous communities, we superimpose them as the squares; the same color indicates that they belong to the same 3D community.
(\textbf{e}) The end-to-end distance for the fractal (FG, blue) and the equilibrium globules (EG, red) averaged over $200$ polymer realizations. The triangles denote the end-to-end distance for community boundaries, and dashed lines represent the chain as a whole. To find the communities, we use $\gamma = 0.4$. The data are obtained from the simulation of $200$ sample globules for each polymer model. The error bars show the standard error of the mean, and the two guided slopes ($1/2$ and $1/3$) show the known scaling of equilibrium and fractal globules at intermediate length scales. 
(\textbf{f}) The same data as in the panel (\textbf{e}), where we scale the vertical axis with the radius of gyration $R_g\left(=\sqrt{\frac{1}{2N^2}\sum_{i,j}\lVert\mathbf{r}_i-\mathbf{r}_j\rVert^2}\right)$ where $N$ is the total number of polymer segments and $\mathbf{r}_i$ is the coordinate of the $i$th segment.}
\label{fig:FG_louvain_heatmap}
\end{figure}

\subsubsection*{Generating fractal globules with the conformation-dependent polymerization algorithm}

As introduced in an article~\cite{Grosberg2016}, there are several variants of fractal globules-like structures~\cite{Rao2014,Sanborn2015,Goloborodko2016,Goloborodko2016a}. Due to simplicity and speed, we use the conformation-dependent polymerization (CDP) model~\cite{Tamm2015}. In a nutshell, this is a Monte Carlo method that produces a fractal globule by simulating a biased random walk on lattice where the propagation probability depends on the entire walk's trajectory over the lattice~\cite{Tamm2015}. This yields on-lattice space-filling polymers. To generate off-lattice fractal globules, the structure is randomized with simulated annealing where the position of a monomer is randomly displaced under the constraint of a fixed inter-monomer distance. With properly chosen parameters, the CDP method produces a fractal globule with contact frequency (probability)  that decays as $\sim s^{-1}$ (as it should~\cite{Mirny2011}), where  $s$ is the contour distance along the polymer. To use the notation from Tamm \emph{et al}.~\cite{Tamm2015}, we use the relation $p \propto (1 + An)$ with $A = 10^4$ for an unoccupied site and $p$ to an occupied site is given by $\epsilon = 10^{-4}$. This result is consistent with the Hi-C data for the human genome at the mega-base-pair scale~\cite{Lieberman-Aiden2009}. We show a realization of a CDP in Fig.~\ref{fig:FG_louvain_heatmap}(\textbf{a}). 

From every simulated CDP polymer, we construct a contact map by counting the number of contact events between polymer beads $i$ and $j$. The contact refers to the case where the Euclidean distance between two beads are shorter than three lattice spacings. To obtain good statistics, we first generate an on-lattice polymer, and then during the annealing stage we register all contacts in $10^3$ random variations of the on-lattice structure. In addition, we have confirmed that different threshold values for what we consider as a contact do not qualitatively alter our results presented below. Just as in the Hi-C experiments, in the final step we normalize the contact map with the KR-norm~\cite{Knight2013} so that each row and column sums to unity.

As a reference case, we use the equilibrium globule. This is a self-avoiding polymer in a closed spherical volume. When we generate equilibrium globules, we contain it in a volume with the same diameter as the fractal globules' radius of gyration. 

\subsubsection*{Structure of 3D communities in fractal globules}
\label{sec:polymer_scailing_properties}

Figure~\ref{fig:FG_louvain_heatmap}(\textbf{b}) shows the contact map for one simulated CDP, where high pixel intensity indicates many contacts. Just as in real Hi-C maps, the simulated map contains locally concentrated contact domains along the diagonal, that is, along the polymer chain.  To find these domains algorithmically, we put the contact map into our modified GenLouvain method. By varying the resolution parameter $\gamma$, we detect communities on various scales. On top of the contact map in Figs.~\ref{fig:FG_louvain_heatmap}(\textbf{b})--(\textbf{d}), we overlay examples of 3D communities for $\gamma=0.4$ (\textbf{b}), $\gamma=0.6$ (\textbf{c}), and $\gamma=0.8$ (\textbf{d}), where the boxes represent community boundaries. Based on these, we ask what the 3D structure of these communities is, and if and how they are different from the polymer as a whole.

Contrasting current views on TADs, we find that 3D communities do not have to be a contiguous polymer segment. Rather, linearly distant parts of the polymer can fold in 3D to form a community. We show this in Fig.~\ref{fig:FG_louvain_heatmap}(\textbf{a}) (left), where we mark the polymer segments belonging to the same community with the same color ($\gamma=0.6$). In Fig.~\ref{fig:FG_louvain_heatmap}(\textbf{a}) (right), we cut out a subsection with three communities and stretch it out. Labeling the communities as A, B, and C, they are clearly ordered in a non-contiguous sequence: they appear as A-B-C-A-B rather than A-B-C.

Furthermore, because of the above-average contact frequencies inside a community, we would like to quantify how its 3D structure differs from the fractal globule polymer as a whole. To do this, we examine the scaling relation of the end-to-end distance---the Euclidean distance between the two boundary monomers defining that (contiguous) community---with respect to the subchain length $s$.  

In Fig.~\ref{fig:FG_louvain_heatmap}(\textbf{e}), we show this relation for the community subchains (the blue triangles) and for all subchains (the blue dashed lines). It shows that the  end-to-end distance for the entire globule grows as $\sim s^{1/3}$, as is expected for a space-filling curve (deviations for large-$s$ comes from  finite-size effects and insufficient statistics). For the communities, we notice that the end-to-end distances are systematically smaller than for a randomly chosen subchain. Our simulations even suggest that the scaling exponent is smaller than $1/3$. Consistent properties are crosschecked in Fig.~\ref{fig:FG_louvain_heatmap}(\textbf{f}) where the end-to-end distance divided by the average chain size ($2\times$radius of gyration) is plotted. Overall, this shows that our method detects 3D communities that are compact substructures of the fractal globule. This observation supports that some TADs in chromosomes are end-closed loop structures~\cite{Rao2014}.
For comparison, we made the same analysis for the equilibrium globule (EG). In Fig.~\ref{fig:FG_louvain_heatmap}(\textbf{e}), we see that the end-to-end distance for our 3D communities and all subchains have the same scaling: $\sim s^{1/2}$ for $s \lesssim N^{2/3}$ and $\sim s^0$ for $s \gtrsim N^{2/3}$, as we expect from an $N$ monomer ideal chain.
Concurrently, the end-to-end distance normalized by the radius of gyration [Fig.~\ref{fig:FG_louvain_heatmap}(\textbf{f})] are almost same for both 3D communities and all subchains, showing a sharp contrast to the FG.

We further investigate the asphericity of the communities' 3D structure in Supplementary Information (SI). We find that the community subchains in FGs tend to be more sphere-like than the FG itself, and the subchains in EGs. This leads to  enhanced contact frequency between monomers within a community (Supplementary Fig.~\ref{fig:AsphericityEffectiveDistance}). Refer to SI for further discussion.

All these analyses support that our community detection method successfully identifies strongly interacting subchains in an FG from the contact map, and that these communities have polymeric properties that cannot be explained by the global expected behavior. Rather, they are more similar to features of TADs. Additionally, in the following section, we corroborate our method by comparing the communities in Hi-C maps detected by our method to the TAD data reported in the literature~\cite{Rao2014}.

\subsection*{Community detection for the Hi-C map}
\label{sec:results}

\begin{figure}[ht]
 \centering
 \begin{tabular}{lll}
  (\textbf{a}) & (\textbf{b}) & (\textbf{c}) \\
 \includegraphics[width=0.25\textwidth]{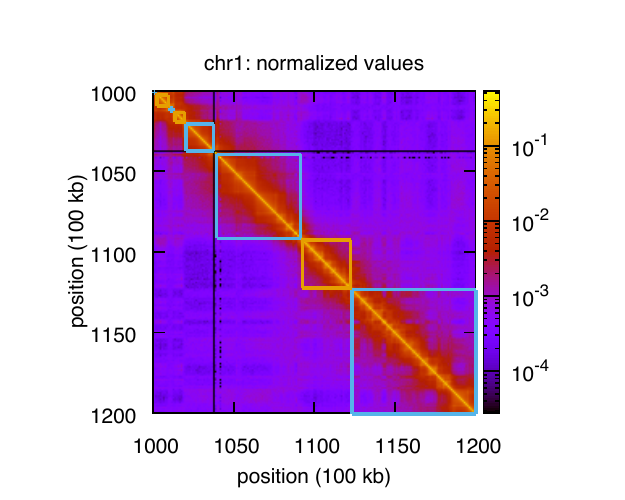} &
 \includegraphics[width=0.25\textwidth]{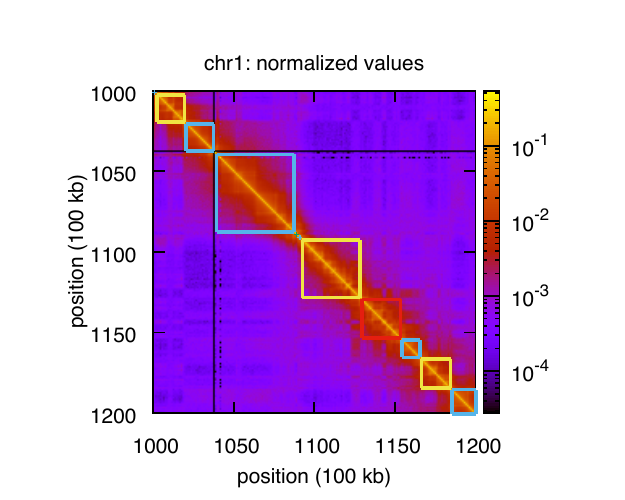} & 
 \includegraphics[width=0.25\textwidth]{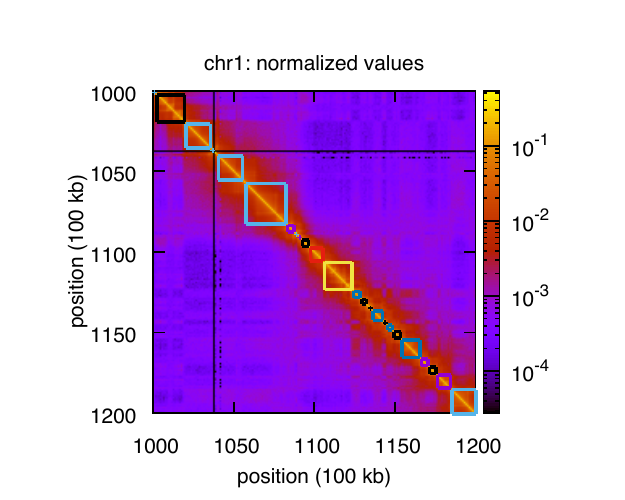} \\
 (\textbf{d}) & (\textbf{e}) & (\textbf{f}) \\
 \includegraphics[width=0.25\textwidth]{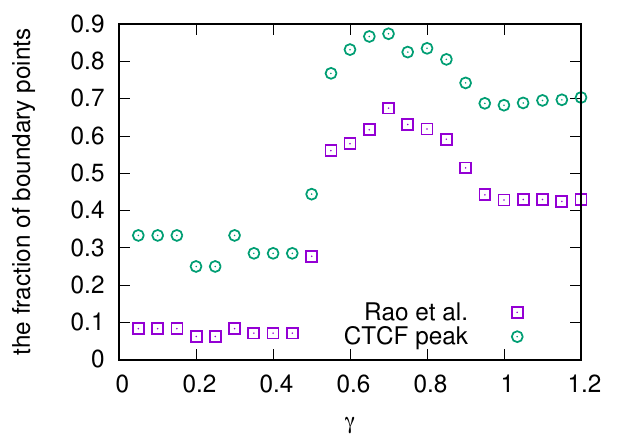} &
 \includegraphics[width=0.3\textwidth]{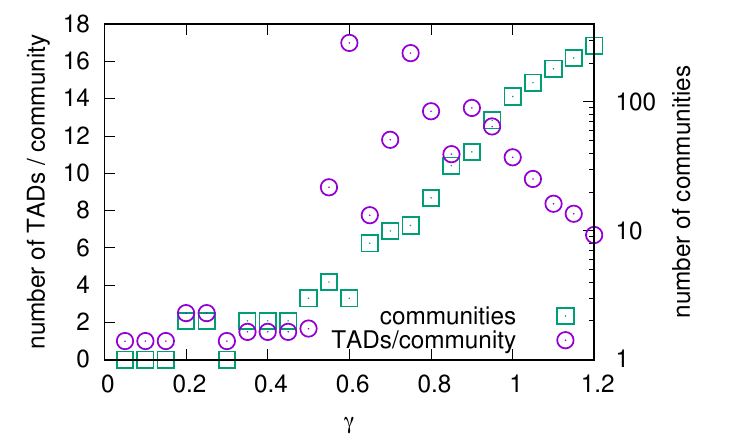} &
 \includegraphics[width=0.2\textwidth]{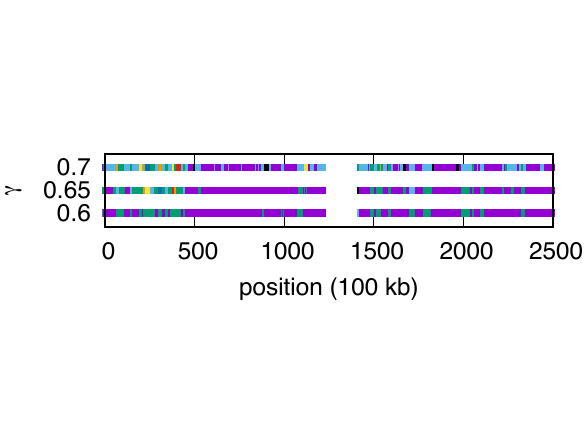} \\
 (\textbf{g}) & (\textbf{h}) & (\textbf{i}) \\
\includegraphics[width=0.25\textwidth]{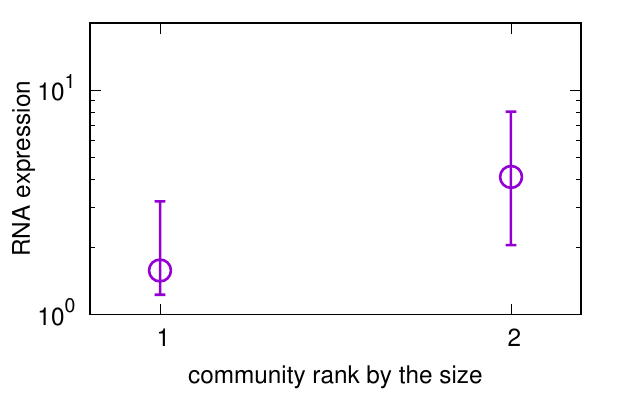} &
\includegraphics[width=0.25\textwidth]{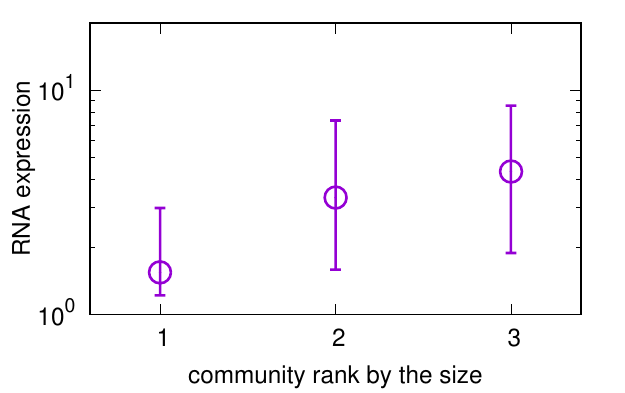} & 
\includegraphics[width=0.25\textwidth]{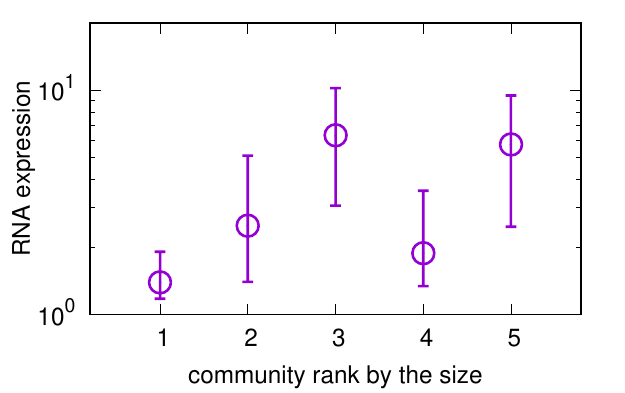} \\
 \end{tabular}
\caption{3D communities in real Hi-C data (chromosome 1). (\textbf{a})--(\textbf{c}): Normalized Hi-C data with squares showing the structure of 3D communities. The black regions are the unmappable regions. The resolution parameter ranges from  $\gamma = 0.6$ (\textbf{a}), $\gamma = 0.7$ (\textbf{b}), and $\gamma = 0.8$ (\textbf{c}). As in Figs.~\ref{fig:FG_louvain_heatmap}(\textbf{b})--(\textbf{d}), we assign the same colors to those squares that belong to the same 3D community. It is clear that they are not contiguous sequences. 
(\textbf{d}) The fraction of community boundary points predicted by our method that coincide with the ones in Rao \emph{et al}.~\cite{Rao2014} (the squares), and binding positions for CTCF (the circles) for different values of $\gamma$.
(\textbf{e}) The average number of TADs for each community, and the number of communities as functions of $\gamma$. 
(\textbf{f}) The community division along chromosome 1 for three different values of $\gamma$. The purple squares represent the largest community in the panels (\textbf{g})--(\textbf{i}), while the other colors indicate smaller communities.
(\textbf{g})--(\textbf{i}) Communities' gene activity sorted by their relative size for different values of $\gamma$: (\textbf{g}) $\gamma = 0.6$, (\textbf{h}) $\gamma = 0.65$, and (\textbf{i}) $\gamma = 0.7$. The circles show the median RNA expression levels, and vertical lines are quartiles. We omit communities that are smaller than $50$ nodes. We find the communities using Eqs.~\eqref{eq:modularity} and \eqref{eq:Q_FG}.
}
\label{fig:real_HiC_results}
 \end{figure}

Based on our community detection approach, we now proceed to analyze Hi-C data from human cells~\cite{Rao2014}. The data comes in the form of matrices where each entry represents the number contacts between two chromosome loci $i$ and $j$. As is standard in the field, we normalized the data with the KR-norm~\cite{Knight2013} which balances the matrix such that every row and column sum to unity (we also used the KR-norm for the fractal globule contact data). The data is available in various resolutions, from $10^3$ base pairs (1 kbp) to $10^6$ base pairs (1 Mbp), but we used 100 kbp which is the scale where both TADs and A/B compartments can be detected~\cite{Rao2014}.

In Figs.~\ref{fig:real_HiC_results}(\textbf{a})--(\textbf{c}), we show that our algorithm detects differently sized contiguous blocks, or TADs, along the chromosome arms as we change the resolution parameter $\gamma$. Similar to the simulated data, it is clear that several TADs form 3D communities. To investigate how these TADs correspond to TADs defined in other studies, we compared the border locations of each contiguous block to TAD borders defined by Rao \emph{et al}.~\cite{Rao2014} (the group that produced the dataset we use in this study) in Fig.~\ref{fig:real_HiC_results}(\textbf{d}). At $\gamma\approx 0.7$, about 70\% of the borders overlap. Moreover, several studies have shown that binding sites for the insulator protein CTCF are strongly correlated with TAD borders (e.g., in Dixon \emph{et al}.~\cite{Dixon2012}). We therefore check the overlap between CTCF binding sites (mapped by ENCODE~\cite{ENCODE}) and all borders at different $\gamma$ values. We note that CTCF has the highest overlap also at $\gamma\approx 0.7$ [Fig.~\ref{fig:real_HiC_results}(\textbf{d})]. 

As we discussed in the introduction, there are different algorithms that detect TADs. However, regardless of the definition that is used, the TADs that come out have substructures that we can interpret as TAD-within-TADs~\cite{weinreb2015identification} with new sets of borders. Our algorithm lets researchers scan through all these TAD-within-TADs. According to Fig.~\ref{fig:real_HiC_results}(\textbf{d}), CTCF correlates well with TAD borders at $\gamma\approx 0.7$. This level also coincides with TADs from Rao \emph{et al}.~\cite{Rao2014}. Our method opens the possibility to investigate which family of proteins is important for different hierarchical levels.

We next investigate properties of the 3D communities. First, we count the number of communities and average number of TADs within each community at different $\gamma$ [Fig.~\ref{fig:real_HiC_results}(\textbf{e})] as well as their localization along the chromosome arms [Fig.~\ref{fig:real_HiC_results}(\textbf{f})]. The highest number of TADs/community is when $\gamma = 0.6 $ and $\gamma = 0.75 $ (about $17$ TADs/community). Interestingly, for $\gamma < 0.6$, the chromosome consists of only two communities (but with several TADs within them). 

Second, we map gene activity within each community using RNA-seq data from ENCODE~\cite{ENCODE} (GEO Acc. Nr: GSE88583). In Figs.~\ref{fig:real_HiC_results}(\textbf{g})--(\textbf{i}) and Supplementary Figs.~\ref{fig:RNAexp_for_communities_chr1_6}--\ref{fig:RNAexp_for_communities_chr19_X}, we show the average coverage of RNA-seq reads within communities for different $\gamma$. At small $\gamma$ values where only two communities are defined, one community is clearly more active than the other. The active and less active communities are then split up into smaller communities as $\gamma$ increases. Already in the original Hi-C paper by Lieberman-Aiden \emph{et al}.~\cite{Lieberman-Aiden2009}, they used principal component analysis on the Hi-C data to partition the chromosomes into two classes. Denoting them by A/B compartments, they found that the A compartment contains transcriptionally active chromatin, whereas the B compartment is less active. 

We took one step in this direction by looking at different types of chromatin. We use 15 chromatin states defined by ENCODE~\cite{Ernst2010,Ernst2011} to see what types of chromatin states are enriched in different communities at different $\gamma$ values (Supplementary Fig.~\ref{fig:comm_state}). These results show that at low values of $\gamma$ one of the two large communities that are identified is enriched in chromatin states associated with transcription, while the other community is not. This again confirms that these represent the A and B compartments respectively. A very small third compartment is also visible that consists of centromere proximal repetitive regions. With increasing $\gamma$ values, we can then see that first the active compartment (A) starts to split up into smaller compartments and then the less active parts of the genome (B) also start to split into smaller communities. It is clear that the genome does not split up into evenly sized sub-compartments, but rather the 3D space is dominated by a few large communities. It is striking and unprecedented that by tuning a single parameter, we detect both TADs and A/B compartments with the same algorithm. We therefore argue that TADs and A/B compartments are not two conceptually different organizational structures in the nucleus, but rather different ends of the same organizational spectrum.

\section*{Discussion}
From Hi-C experiments, it is clear that inter-phase chromosomes are built up by a network of 3D compartments on various scales---from kilo($10^3$)-base-pair sized loops to mega($10^6$)-base-pair sized 3D structures. This pattern is consistent across organisms and cell types. To let researchers scan through the spectrum of 3D compartments, we have tailored the GenLouvain community detection method to find 3D communities in fractal globule polymer systems. Apart from verifying our method on computer-generated polymers, we have applied it to analyze human Hi-C data. First, we have found that chromatin segments belonging to the same 3D community do not have to be in next to each other along the DNA. In other words, several TADs can belong to the same 3D community. Second, we have found that CTCF proteins---a loop-stabilizing protein that is ascribed a big role in TAD formation---are only correlated well with community borders at one level of organization. It remains to find what other factors are important at higher or lower levels. Third, just by adjusting a single parameter ($\gamma$), our method picks up the two most prominent 3D compartments, TADs and A/B compartments, which are traditionally treated as two weakly related 3D structures and detected with different algorithms. Rather than seeing them as different, our work put them on an equal footing, and we argue that they represent two ends of a continuous spectrum of 3D communities of different sizes. 

\bibliography{main}

\begin{thebibliography}{10}
\urlstyle{rm}
\expandafter\ifx\csname url\endcsname\relax
  \def\url#1{\texttt{#1}}\fi
\expandafter\ifx\csname urlprefix\endcsname\relax\def\urlprefix{URL }\fi
\expandafter\ifx\csname doiprefix\endcsname\relax\def\doiprefix{DOI: }\fi
\providecommand{\bibinfo}[2]{#2}
\providecommand{\eprint}[2][]{\url{#2}}

\bibitem{Lieberman-Aiden2009}
\bibinfo{author}{Lieberman-Aiden, E.} \emph{et~al.}
\newblock \bibinfo{journal}{\bibinfo{title}{Comprehensive mapping of long-range
  interactions reveals folding principles of the human genome}}.
\newblock {\emph{\JournalTitle{Science}}} \textbf{\bibinfo{volume}{326}},
  \bibinfo{pages}{289--293}, \doiprefix\url{10.1126/science.1181369}
  (\bibinfo{year}{2009}).
\newblock \eprint{http://science.sciencemag.org/content/326/5950/289.full.pdf}.

\bibitem{Rao2014}
\bibinfo{author}{Rao, S.~S.} \emph{et~al.}
\newblock \bibinfo{journal}{\bibinfo{title}{{{A} 3{D} map of the human genome
  at kilobase resolution reveals principles of chromatin looping}}}.
\newblock {\emph{\JournalTitle{Cell}}} \textbf{\bibinfo{volume}{159}},
  \bibinfo{pages}{1665--1680} (\bibinfo{year}{2014}).

\bibitem{Dixon2012}
\bibinfo{author}{Dixon, J.~R.} \emph{et~al.}
\newblock \bibinfo{journal}{\bibinfo{title}{Topological domains in mammalian
  genomes identified by analysis of chromatin interactions}}.
\newblock {\emph{\JournalTitle{Nature}}} \textbf{\bibinfo{volume}{485}},
  \bibinfo{pages}{376--380} (\bibinfo{year}{2012}).

\bibitem{Nora2012}
\bibinfo{author}{Nora, E.~P.} \emph{et~al.}
\newblock \bibinfo{journal}{\bibinfo{title}{Spatial partitioning of the
  regulatory landscape of the x-inactivation centre}}.
\newblock {\emph{\JournalTitle{Nature}}} \textbf{\bibinfo{volume}{485}},
  \bibinfo{pages}{381--385} (\bibinfo{year}{2012}).

\bibitem{dixon2015chromatin}
\bibinfo{author}{Dixon, J.~R.} \emph{et~al.}
\newblock \bibinfo{journal}{\bibinfo{title}{Chromatin architecture
  reorganization during stem cell differentiation}}.
\newblock {\emph{\JournalTitle{Nature}}} \textbf{\bibinfo{volume}{518}},
  \bibinfo{pages}{331--336} (\bibinfo{year}{2015}).

\bibitem{Boulos2013}
\bibinfo{author}{Boulos, R.~E.}, \bibinfo{author}{Arneodo, A.},
  \bibinfo{author}{Jensen, P.} \& \bibinfo{author}{Audit, B.}
\newblock \bibinfo{journal}{\bibinfo{title}{Revealing long-range interconnected
  hubs in human chromatin interaction data using graph theory}}.
\newblock {\emph{\JournalTitle{Phys. Rev. Lett.}}}
  \textbf{\bibinfo{volume}{111}}, \bibinfo{pages}{118102},
  \doiprefix\url{10.1103/PhysRevLett.111.118102} (\bibinfo{year}{2013}).

\bibitem{Cabreros2015}
\bibinfo{author}{Cabreros, I.}, \bibinfo{author}{Abbe, E.} \&
  \bibinfo{author}{Tsirigos, A.}
\newblock \bibinfo{title}{Detecting community structures in {Hi-C} genomic
  data}.
\newblock In \emph{\bibinfo{booktitle}{2016 Annual Conference on Information
  Science and Systems (CISS)}}, \bibinfo{pages}{584--589},
  \doiprefix\url{10.1109/CISS.2016.7460568} (\bibinfo{year}{2016}).

\bibitem{YXRWang2017}
\bibinfo{author}{{Wang}, Y.~X.~R.}, \bibinfo{author}{{Sarkar}, P.},
  \bibinfo{author}{{Ursu}, O.}, \bibinfo{author}{{Kundaje}, A.} \&
  \bibinfo{author}{{Bickel}, P.~J.}
\newblock \bibinfo{journal}{\bibinfo{title}{{Network modelling of topological
  domains using Hi-C data}}}.
\newblock {\emph{\JournalTitle{e-print arXiv:1707.09587}}}
  (\bibinfo{year}{2017}).

\bibitem{Sarnataro2017}
\bibinfo{author}{Sarnataro, S.}, \bibinfo{author}{Chiariello, A.~M.},
  \bibinfo{author}{Esposito, A.}, \bibinfo{author}{Prisco, A.} \&
  \bibinfo{author}{Nicodemi, M.}
\newblock \bibinfo{journal}{\bibinfo{title}{{{S}tructure of the human
  chromosome interaction network}}}.
\newblock {\emph{\JournalTitle{PLoS ONE}}} \textbf{\bibinfo{volume}{12}},
  \bibinfo{pages}{e0188201} (\bibinfo{year}{2017}).

\bibitem{Belyaeva2017}
\bibinfo{author}{Belyaeva, A.}, \bibinfo{author}{Venkatachalapathy, S.},
  \bibinfo{author}{Nagarajan, M.}, \bibinfo{author}{Shivashankar, G.~V.} \&
  \bibinfo{author}{Uhler, C.}
\newblock \bibinfo{journal}{\bibinfo{title}{Network analysis identifies
  chromosome intermingling regions as regulatory hotspots for transcription}}.
\newblock {\emph{\JournalTitle{Proceedings of the National Academy of
  Sciences}}} \textbf{\bibinfo{volume}{114}}, \bibinfo{pages}{13714--13719},
  \doiprefix\url{10.1073/pnas.1708028115} (\bibinfo{year}{2017}).
\newblock \eprint{http://www.pnas.org/content/114/52/13714.full.pdf}.

\bibitem{yu2017identifying}
\bibinfo{author}{Yu, W.}, \bibinfo{author}{He, B.} \& \bibinfo{author}{Tan, K.}
\newblock \bibinfo{journal}{\bibinfo{title}{Identifying topologically
  associating domains and subdomains by gaussian mixture model and proportion
  test}}.
\newblock {\emph{\JournalTitle{Nature Communications}}}
  \textbf{\bibinfo{volume}{8}}, \bibinfo{pages}{535},
  \doiprefix\url{10.1038/s41467-017-00478-8} (\bibinfo{year}{2017}).

\bibitem{weinreb2015identification}
\bibinfo{author}{Weinreb, C.} \& \bibinfo{author}{Raphael, B.~J.}
\newblock \bibinfo{journal}{\bibinfo{title}{Identification of hierarchical
  chromatin domains}}.
\newblock {\emph{\JournalTitle{Bioinformatics}}} \textbf{\bibinfo{volume}{32}},
  \bibinfo{pages}{1601--1609}, \doiprefix\url{10.1093/bioinformatics/btv485}
  (\bibinfo{year}{2016}).

\bibitem{haddad2017ic}
\bibinfo{author}{Haddad, N.}, \bibinfo{author}{Vaillant, C.} \&
  \bibinfo{author}{Jost, D.}
\newblock \bibinfo{journal}{\bibinfo{title}{Ic-finder: inferring robustly the
  hierarchical organization of chromatin folding}}.
\newblock {\emph{\JournalTitle{Nucleic Acids Research}}}
  \textbf{\bibinfo{volume}{45}}, \bibinfo{pages}{e81},
  \doiprefix\url{10.1093/nar/gkx036} (\bibinfo{year}{2017}).

\bibitem{KKYan2017}
\bibinfo{author}{Yan, K.~K.}, \bibinfo{author}{Lou, S.} \&
  \bibinfo{author}{Gerstein, M.}
\newblock \bibinfo{journal}{\bibinfo{title}{{{M}r{T}{A}{D}{F}inder: {A} network
  modularity based approach to identify topologically associating domains in
  multiple resolutions}}}.
\newblock {\emph{\JournalTitle{PLoS Comput. Biol.}}}
  \textbf{\bibinfo{volume}{13}}, \bibinfo{pages}{e1005647}
  (\bibinfo{year}{2017}).

\bibitem{Norton2018}
\bibinfo{author}{Norton, H.~K.} \emph{et~al.}
\newblock \bibinfo{journal}{\bibinfo{title}{Detecting hierarchical genome
  folding with network modularity}}.
\newblock {\emph{\JournalTitle{Nature Methods}}} \textbf{\bibinfo{volume}{15}},
  \bibinfo{pages}{119--122} (\bibinfo{year}{2018}).

\bibitem{Ball2011}
\bibinfo{author}{Ball, B.}, \bibinfo{author}{Karrer, B.} \&
  \bibinfo{author}{Newman, M. E.~J.}
\newblock \bibinfo{journal}{\bibinfo{title}{Efficient and principled method for
  detecting communities in networks}}.
\newblock {\emph{\JournalTitle{Phys. Rev. E}}} \textbf{\bibinfo{volume}{84}},
  \bibinfo{pages}{036103}, \doiprefix\url{10.1103/PhysRevE.84.036103}
  (\bibinfo{year}{2011}).

\bibitem{Gopalan2013}
\bibinfo{author}{Gopalan, P.~K.} \& \bibinfo{author}{Blei, D.~M.}
\newblock \bibinfo{journal}{\bibinfo{title}{Efficient discovery of overlapping
  communities in massive networks}}.
\newblock {\emph{\JournalTitle{Proceedings of the National Academy of
  Sciences}}} \textbf{\bibinfo{volume}{110}}, \bibinfo{pages}{14534--14539},
  \doiprefix\url{10.1073/pnas.1221839110} (\bibinfo{year}{2013}).
\newblock \eprint{http://www.pnas.org/content/110/36/14534.full.pdf}.

\bibitem{Newman2016}
\bibinfo{author}{Newman, M. E.~J.}
\newblock \bibinfo{journal}{\bibinfo{title}{Equivalence between modularity
  optimization and maximum likelihood methods for community detection}}.
\newblock {\emph{\JournalTitle{Phys. Rev. E}}} \textbf{\bibinfo{volume}{94}},
  \bibinfo{pages}{052315}, \doiprefix\url{10.1103/PhysRevE.94.052315}
  (\bibinfo{year}{2016}).

\bibitem{Porter2009}
\bibinfo{author}{Porter, M.~A.}, \bibinfo{author}{Onnela, J.~P.} \&
  \bibinfo{author}{Mucha, P.~J.}
\newblock \bibinfo{journal}{\bibinfo{title}{Communities in networks}}.
\newblock {\emph{\JournalTitle{Not. Am. Math. Soc.}}}
  \textbf{\bibinfo{volume}{56}}, \bibinfo{pages}{1082--1092, 1164--1166}
  (\bibinfo{year}{2009}).

\bibitem{Fortunato2010}
\bibinfo{author}{Fortunato, S.}
\newblock \bibinfo{journal}{\bibinfo{title}{Community detection in graphs}}.
\newblock {\emph{\JournalTitle{Physics Reports}}}
  \textbf{\bibinfo{volume}{486}}, \bibinfo{pages}{75--174},
  \doiprefix\url{https://doi.org/10.1016/j.physrep.2009.11.002}
  (\bibinfo{year}{2010}).

\bibitem{GenLouvain}
\bibinfo{author}{Jutla, I.~S.}, \bibinfo{author}{Jeub, L. G.~S.} \&
  \bibinfo{author}{Mucha, P.~J.}
\newblock \bibinfo{title}{Generalized {Louvain} method for community detection
  implemented in {MATLAB} ({GenLouvain} 2.0)}.
\newblock
  \bibinfo{howpublished}{\url{http://netwiki.amath.unc.edu/GenLouvain/GenLouvain}}
  (\bibinfo{year}{2010--2014}).

\bibitem{Blondel2008}
\bibinfo{author}{Blondel, V.~D.}, \bibinfo{author}{Guillaume, J.-L.},
  \bibinfo{author}{Lambiotte, R.} \& \bibinfo{author}{Lefebvre, E.}
\newblock \bibinfo{journal}{\bibinfo{title}{Fast unfolding of communities in
  large networks}}.
\newblock {\emph{\JournalTitle{Journal of Statistical Mechanics: Theory and
  Experiment}}} \textbf{\bibinfo{volume}{2008}}, \bibinfo{pages}{P10008}
  (\bibinfo{year}{2008}).

\bibitem{Fraser2015}
\bibinfo{author}{Fraser, J.} \emph{et~al.}
\newblock \bibinfo{journal}{\bibinfo{title}{Hierarchical folding and
  reorganization of chromosomes are linked to transcriptional changes~in
  cellular differentiation}}.
\newblock {\emph{\JournalTitle{Molecular Systems Biology}}}
  \textbf{\bibinfo{volume}{11}}, \bibinfo{pages}{852},
  \doiprefix\url{10.15252/msb.20156492} (\bibinfo{year}{2015}).
\newblock \eprint{http://msb.embopress.org/content/11/12/852.full.pdf}.

\bibitem{Bianco2017}
\bibinfo{author}{Bianco, S.}, \bibinfo{author}{Chiariello, A.~M.},
  \bibinfo{author}{Annunziatella, C.}, \bibinfo{author}{Esposito, A.} \&
  \bibinfo{author}{Nicodemi, M.}
\newblock \bibinfo{journal}{\bibinfo{title}{Predicting chromatin architecture
  from models of polymer physics}}.
\newblock {\emph{\JournalTitle{Chromosome Research}}}
  \textbf{\bibinfo{volume}{25}}, \bibinfo{pages}{25--34},
  \doiprefix\url{10.1007/s10577-016-9545-5} (\bibinfo{year}{2017}).

\bibitem{Newman2004}
\bibinfo{author}{Newman, M. E.~J.} \& \bibinfo{author}{Girvan, M.}
\newblock \bibinfo{journal}{\bibinfo{title}{Finding and evaluating community
  structure in networks}}.
\newblock {\emph{\JournalTitle{Phys. Rev. E}}} \textbf{\bibinfo{volume}{69}},
  \bibinfo{pages}{026113}, \doiprefix\url{10.1103/PhysRevE.69.026113}
  (\bibinfo{year}{2004}).

\bibitem{burton2014species}
\bibinfo{author}{Burton, J.~N.}, \bibinfo{author}{Liachko, I.},
  \bibinfo{author}{Dunham, M.~J.} \& \bibinfo{author}{Shendure, J.}
\newblock \bibinfo{journal}{\bibinfo{title}{Species-level deconvolution of
  metagenome assemblies with {Hi-C}{\textendash}based contact probability
  maps}}.
\newblock {\emph{\JournalTitle{G3: Genes, Genomes, Genetics}}}
  \textbf{\bibinfo{volume}{4}}, \bibinfo{pages}{1339--1346},
  \doiprefix\url{10.1534/g3.114.011825} (\bibinfo{year}{2014}).
\newblock \eprint{http://www.g3journal.org/content/4/7/1339.full.pdf}.

\bibitem{yu2012spatial}
\bibinfo{author}{Zhang, Y.} \emph{et~al.}
\newblock \bibinfo{journal}{\bibinfo{title}{{{S}patial organization of the
  mouse genome and its role in recurrent chromosomal translocations}}}.
\newblock {\emph{\JournalTitle{Cell}}} \textbf{\bibinfo{volume}{148}},
  \bibinfo{pages}{908--921} (\bibinfo{year}{2012}).

\bibitem{Tamm2015}
\bibinfo{author}{Tamm, M.~V.}, \bibinfo{author}{Nazarov, L.~I.},
  \bibinfo{author}{Gavrilov, A.~A.} \& \bibinfo{author}{Chertovich, A.~V.}
\newblock \bibinfo{journal}{\bibinfo{title}{Anomalous diffusion in fractal
  globules}}.
\newblock {\emph{\JournalTitle{Phys. Rev. Lett.}}}
  \textbf{\bibinfo{volume}{114}}, \bibinfo{pages}{178102},
  \doiprefix\url{10.1103/PhysRevLett.114.178102} (\bibinfo{year}{2015}).

\bibitem{Mirny2011}
\bibinfo{author}{Mirny, L.~A.}
\newblock \bibinfo{journal}{\bibinfo{title}{The fractal globule as a model of
  chromatin architecture in the cell}}.
\newblock {\emph{\JournalTitle{Chromosome Research}}}
  \textbf{\bibinfo{volume}{19}}, \bibinfo{pages}{37--51},
  \doiprefix\url{10.1007/s10577-010-9177-0} (\bibinfo{year}{2011}).

\bibitem{grosberg1993crumpled}
\bibinfo{author}{Grosberg, A.}, \bibinfo{author}{Rabin, Y.},
  \bibinfo{author}{Havlin, S.} \& \bibinfo{author}{Neer, A.}
\newblock \bibinfo{journal}{\bibinfo{title}{Crumpled globule model of the
  three-dimensional structure of {DNA}}}.
\newblock {\emph{\JournalTitle{EPL}}} \textbf{\bibinfo{volume}{23}},
  \bibinfo{pages}{373} (\bibinfo{year}{1993}).

\bibitem{Grosberg2016}
\bibinfo{author}{Grosberg, A.~Y.}
\newblock \bibinfo{journal}{\bibinfo{title}{{Extruding loops to make loopy
  globules?}}}
\newblock {\emph{\JournalTitle{Biophys. J.}}} \textbf{\bibinfo{volume}{110}},
  \bibinfo{pages}{2133--2135} (\bibinfo{year}{2016}).

\bibitem{Sanborn2015}
\bibinfo{author}{Sanborn, A.~L.} \emph{et~al.}
\newblock \bibinfo{journal}{\bibinfo{title}{Chromatin extrusion explains key
  features of loop and domain formation in wild-type and engineered genomes}}.
\newblock {\emph{\JournalTitle{Proceedings of the National Academy of
  Sciences}}} \textbf{\bibinfo{volume}{112}}, \bibinfo{pages}{E6456--E6465},
  \doiprefix\url{10.1073/pnas.1518552112} (\bibinfo{year}{2015}).
\newblock \eprint{https://www.pnas.org/content/112/47/E6456.full.pdf}.

\bibitem{Goloborodko2016}
\bibinfo{author}{Goloborodko, A.}, \bibinfo{author}{Marko, J.~F.} \&
  \bibinfo{author}{Mirny, L.~A.}
\newblock \bibinfo{journal}{\bibinfo{title}{{{C}hromosome compaction by active
  loop extrusion}}}.
\newblock {\emph{\JournalTitle{Biophys. J.}}} \textbf{\bibinfo{volume}{110}},
  \bibinfo{pages}{2162--2168} (\bibinfo{year}{2016}).

\bibitem{Goloborodko2016a}
\bibinfo{author}{Goloborodko, A.}, \bibinfo{author}{Imakaev, M.~V.},
  \bibinfo{author}{Marko, J.~F.} \& \bibinfo{author}{Mirny, L.}
\newblock \bibinfo{journal}{\bibinfo{title}{{{C}ompaction and segregation of
  sister chromatids via active loop extrusion}}}.
\newblock {\emph{\JournalTitle{Elife}}} \textbf{\bibinfo{volume}{5}}
  (\bibinfo{year}{2016}).

\bibitem{Knight2013}
\bibinfo{author}{Knight, P.~A.} \& \bibinfo{author}{Ruiz, D.}
\newblock \bibinfo{journal}{\bibinfo{title}{A fast algorithm for matrix
  balancing}}.
\newblock {\emph{\JournalTitle{IMA Journal of Numerical Analysis}}}
  \textbf{\bibinfo{volume}{33}}, \bibinfo{pages}{1029--1047},
  \doiprefix\url{10.1093/imanum/drs019} (\bibinfo{year}{2013}).
\newblock
  \eprint{/oup/backfile/content_public/journal/imajna/33/3/10.1093/imanum/drs019/2/drs019.pdf}.

\bibitem{ENCODE}
\bibinfo{author}{Dunham, I.} \emph{et~al.}
\newblock \bibinfo{journal}{\bibinfo{title}{An integrated encyclopedia of dna
  elements in the human genome}}.
\newblock {\emph{\JournalTitle{Nature}}} \textbf{\bibinfo{volume}{489}},
  \bibinfo{pages}{57--74} (\bibinfo{year}{2012}).

\bibitem{Ernst2010}
\bibinfo{author}{Ernst, J.} \& \bibinfo{author}{Kellis, M.}
\newblock \bibinfo{journal}{\bibinfo{title}{{{D}iscovery and characterization
  of chromatin states for systematic annotation of the human genome}}}.
\newblock {\emph{\JournalTitle{Nat. Biotechnol.}}}
  \textbf{\bibinfo{volume}{28}}, \bibinfo{pages}{817--825}
  (\bibinfo{year}{2010}).

\bibitem{Ernst2011}
\bibinfo{author}{Ernst, J.} \emph{et~al.}
\newblock \bibinfo{journal}{\bibinfo{title}{{{M}apping and analysis of
  chromatin state dynamics in nine human cell types}}}.
\newblock {\emph{\JournalTitle{Nature}}} \textbf{\bibinfo{volume}{473}},
  \bibinfo{pages}{43--49} (\bibinfo{year}{2011}).

\bibitem{rudnick1987shapes}
\bibinfo{author}{Rudnick, J.} \& \bibinfo{author}{Gaspari, G.}
\newblock \bibinfo{journal}{\bibinfo{title}{The shapes of random walks}}.
\newblock {\emph{\JournalTitle{Science}}} \textbf{\bibinfo{volume}{237}},
  \bibinfo{pages}{384--389}, \doiprefix\url{10.1126/science.237.4813.384}
  (\bibinfo{year}{1987}).
\newblock \eprint{https://doi.org/10.1126/science.237.4813.384}.

\end{thebibliography}

\section*{Acknowledgements}
The authors thank Rajendra Kumar for fruitful discussion and the practical help in various data processing throughout our work. This work was supported by the National Research Foundation (NRF) of Korea (No. 2016K2A9A2A12003797 and No. 2017K1A1A2013241), and the Knut and Alice Wallenberg foundation (grant number 2014-0018, to EpiCoN, co-PI: PS). S.H.L. was also supported by the NRF of Korea (No. 2018R1C1B5083863). S.L. was supported by Basic Science Research Program
through the NRF of Korea funded by the Ministry of Education (No. 2016R1A6A3A11932833). L.L. was also supported by the Swedish Foundation for International Cooperation in Research and Higher Education (STINT) (No. KO2015-6452).

\section*{Author contributions}
S.H.L., S.L., X.D., P.S., J.-H.J., and L.L. designed research, S.H.L., Y.K., and P.S. conducted the data processing and numerical simulations, S.H.L., Y.K., S.L., P.S., J.-H.J, and L.L. analyzed the results. All authors wrote and reviewed the manuscript. 

\section*{Additional Information}

\textbf{Competing Interests}: The authors declare no competing interests.. 





\clearpage

\renewcommand{\figurename}{Supplementary Figure}
\renewcommand{\thefigure}{S\arabic{figure}}
\setcounter{figure}{0}

\section*{Supplementary Information}
In this Supplementary Information, we present additional investigations on the polymeric properties of the community subchains in FGs.

\subsection*{Asphericity of the 3D folded structure of community subchains}
The asymmetric 3D shape of a polymer chain can be analyzed using the gyration tensor $\mathbf{S}$. For a polymer chain composed of $N$ monomers, the element of the gyration tensor is defined as
\begin{equation}
S_{mn} = \frac{1}{2N^2} \sum_{i=1}^N \sum_{j=1}^N \left( r_m^{(i)} - r_m^{(j)} \right) \left( r_n^{(i)} - r_n^{(j)} \right) \,,
\label{eq:gyration_tensor}
\end{equation}
where $r_n^{(i)}$ is the $n$th Cartesian coordinate of the monomer $i \in \{1, \cdots, N\}$. The three eigenvalues of the gyration tensor in Eq.~\eqref{eq:gyration_tensor}, denoted by $\lambda_1 ^2$, $\lambda_2 ^2$, and $\lambda_3 ^2$, characterize the overall geometry of the polymer. The radius of gyration, defined in Fig.~\ref{fig:FG_louvain_heatmap}(\textbf{f}) in the main text, is obtained as $R_g^2=S_{xx}+S_{yy}+S_{zz}$ and, using the eigenvalues, given by
\begin{equation}
R_g = \sqrt{\lambda_1^2 + \lambda_2^2 + \lambda_3^2} \,.
\label{eq:radius_of_gyration}
\end{equation}

The asphericity measuring the relative shape anisotropy is quantified by the parameter $\kappa^2$, defined as
\begin{equation}
\kappa^2 = \frac{\left(\lambda_1^2-\lambda_2^2\right)^2 + \left(\lambda_2^2-\lambda_3^2\right)^2 + \left(\lambda_3^2-\lambda_1^2\right)^2}{2\left(\lambda_1^2 + \lambda_2^2 + \lambda_3^2\right)^2} = \frac{3\left(\lambda_1^4 + \lambda_2^4 + \lambda_3^4 \right)}{2 \left( \lambda_1^2 + \lambda_2^2 + \lambda_3^2  \right)^2} - \frac{1}{2} \,.
\label{eq:asphericity}
\end{equation}
$\kappa^2$ varies from 0 (isotropic objects having the uniform eigenvalues) to 1 (severely elongated objects having a single dominant eigenvalue). In Supplementary Fig.~\ref{fig:AsphericityEffectiveDistance}, we plot the asphericity $\kappa^2$ as a function of subchain length for the communities and randomly chosen subchains in fractal and equilibrium globules. For an ideal unconfined chain (dashed-dotted line), the asphericity is known to be $\kappa^2\simeq 0.39$~\cite{rudnick1987shapes}. When for FGs and EGs the subchain length is too short to feel the confinement, $\kappa^2$ for the subchains has roughly this ideal value, and decreases with the chain length. We find that, compared to the average subchains in FGs, the communities therein have smaller $\kappa^2$ values, thus being more sphere-like than the FG itself. Together with the end-to-end distance analysis in Fig.\ref{fig:FG_louvain_heatmap} in the main text, this indicates that the community subchains in FGs have a 3D structure where both ends (i.e., the boundary sites in its contact map) are close to each other and the monomer distribution is almost isotropic. The reference system (EGs) shows a similar trend while the communities in FGs are more sphere-like than in EGs up to the subchain length $\sim 500$. Beyond this length, $\kappa^2$ for EGs approaches zero for the randomly chosen subchains, due to the finite-size effect caused by the simulation protocol for making an EG using the random walk inside a sphere.

\subsection*{The results for all of the chromosomes}
\begin{figure}
\centering
\begin{tabular}{cc}
chromosome 7 & chromosome 8 \\
\includegraphics[width=0.4\textwidth]{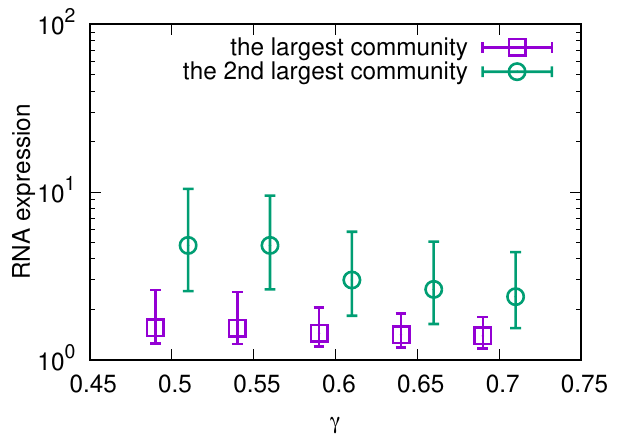} &
\includegraphics[width=0.4\textwidth]{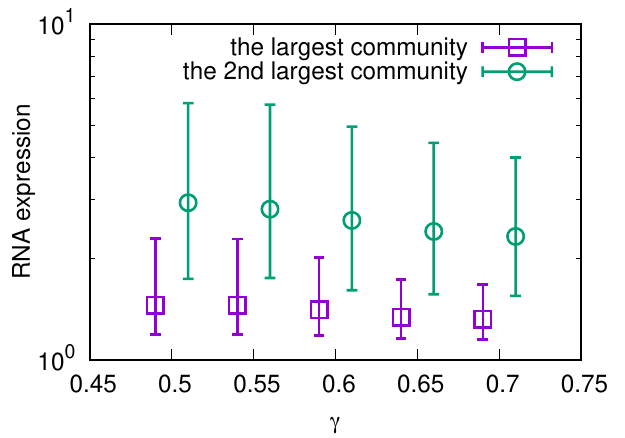} \\
chromosome 9 & chromosome 10 \\
\includegraphics[width=0.4\textwidth]{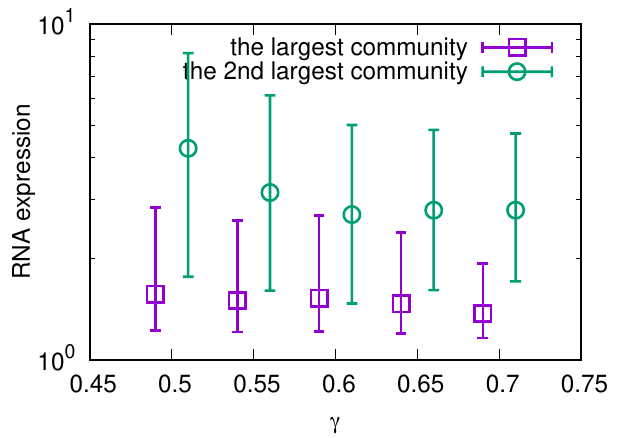} &
\includegraphics[width=0.4\textwidth]{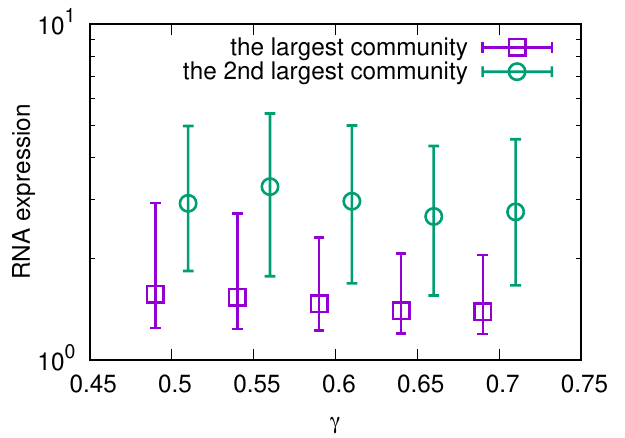} \\
chromosome 11 & chromosome 12 \\
\includegraphics[width=0.4\textwidth]{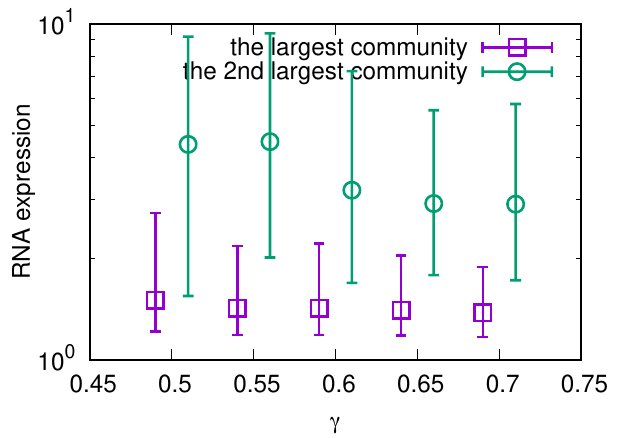} &
\includegraphics[width=0.4\textwidth]{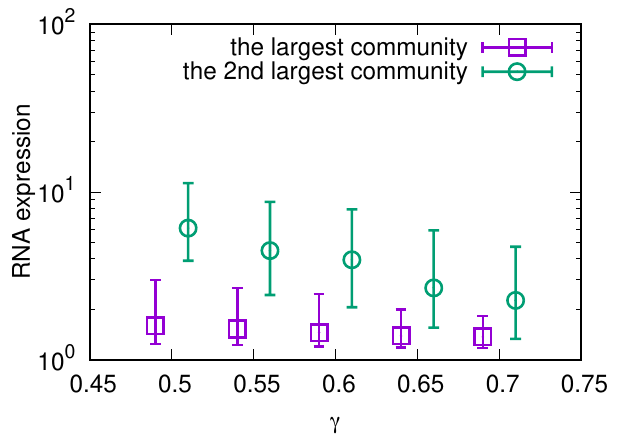} \\
\end{tabular}
\caption{Differential RNA expression levels for different communities. We present the differential RNA expression for communities corresponding to various values of $\gamma$ for chromosomes $7$--$12$, only for the largest and second largest communities. For visualization, we shift the data points slightly to the left for the largest community and right for the second largest community for each of the $\gamma$ values. The plots indicate the median values with the quartiles as the error bars.
}
\label{fig:RNAexp_for_communities_chr7_12}
\end{figure}

\begin{figure}
\centering
\begin{tabular}{cc}
chromosome 13 & chromosome 14 \\
\includegraphics[width=0.4\textwidth]{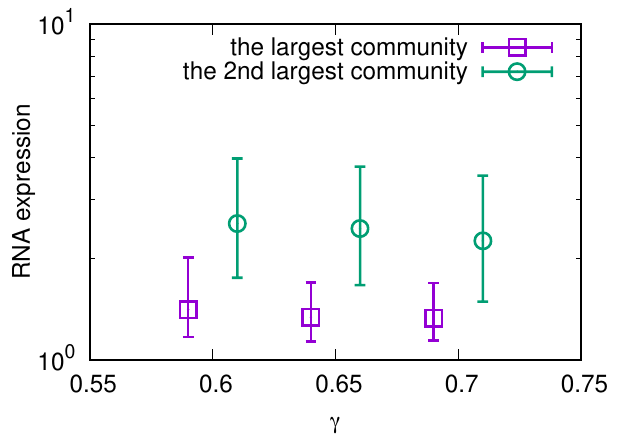} &
\includegraphics[width=0.4\textwidth]{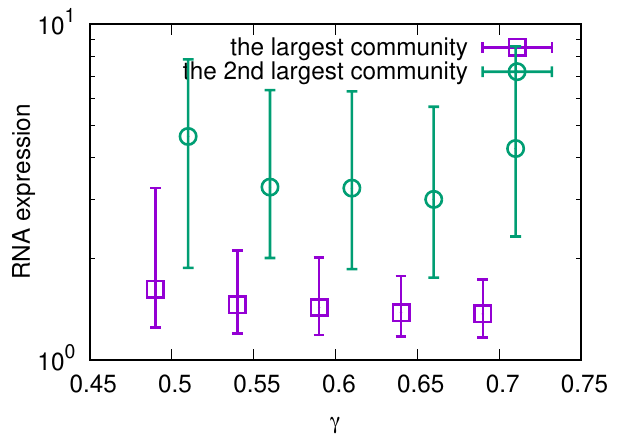} \\
chromosome 15 & chromosome 16 \\
\includegraphics[width=0.4\textwidth]{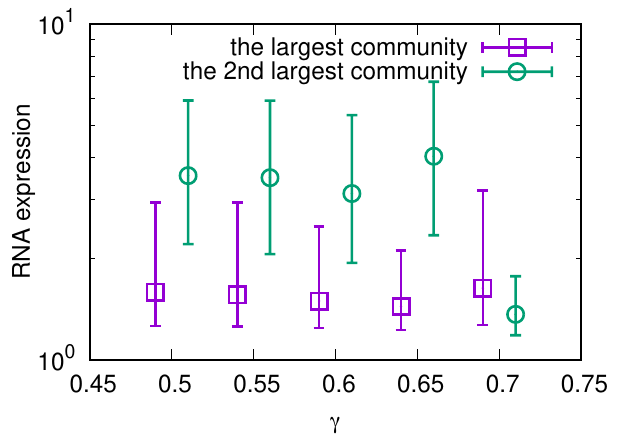} &
\includegraphics[width=0.4\textwidth]{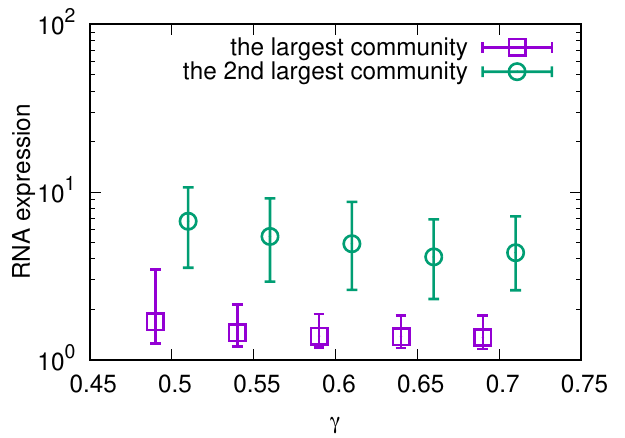} \\
chromosome 17 & chromosome 18 \\
\includegraphics[width=0.4\textwidth]{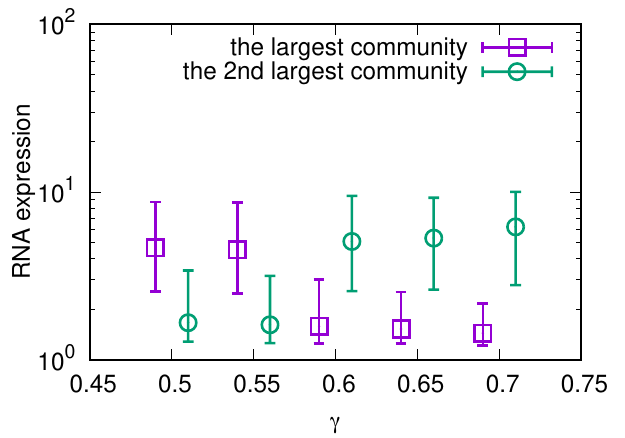} &
\includegraphics[width=0.4\textwidth]{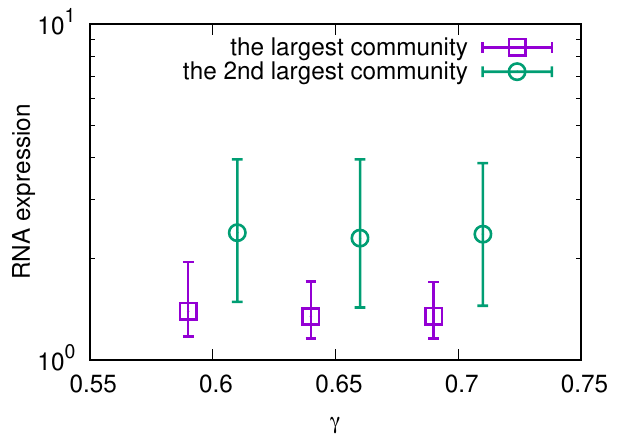} \\
\end{tabular}
\caption{Differential RNA expression levels for different communities. We present the differential RNA expression for communities corresponding to various values of $\gamma$ for chromosomes $13$--$18$, only for the largest and second largest communities. For visualization, we shift the data points slightly to the left for the largest community and right for the second largest community for each of the $\gamma$ values. The plots indicate the median values with the quartiles as the error bars.
}
\label{fig:RNAexp_for_communities_chr13_18}
\end{figure}

\begin{figure}
\centering
\begin{tabular}{cc}
chromosome 19 & chromosome 20 \\
\includegraphics[width=0.4\textwidth]{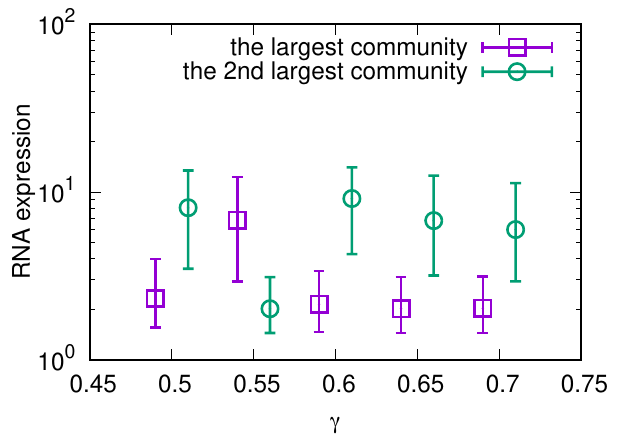} &
\includegraphics[width=0.4\textwidth]{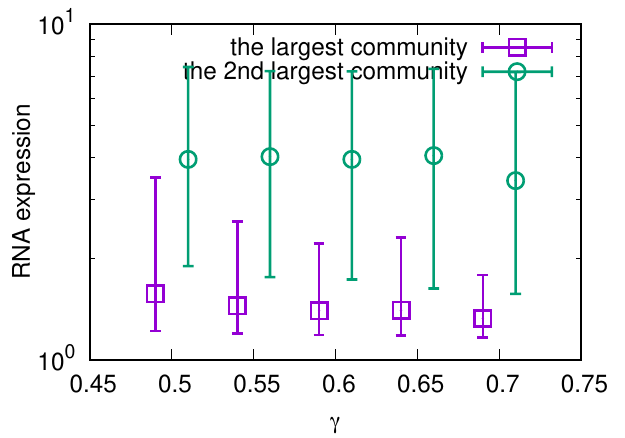} \\
chromosome 21 & chromosome 22 \\
\includegraphics[width=0.4\textwidth]{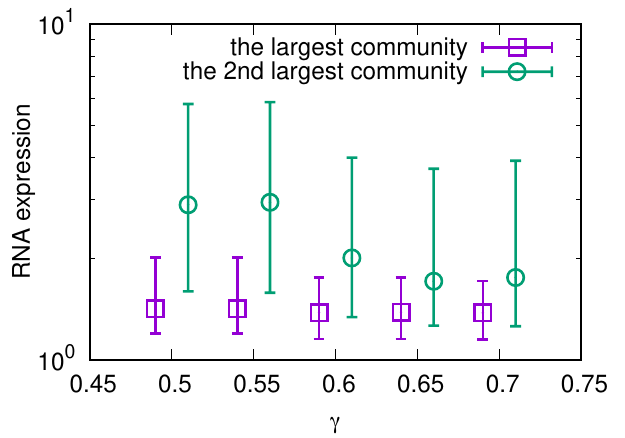} &
\includegraphics[width=0.4\textwidth]{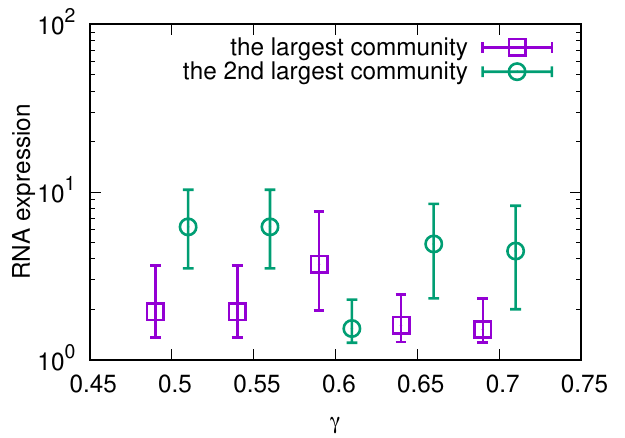} \\
chromosome X &  \\
\includegraphics[width=0.4\textwidth]{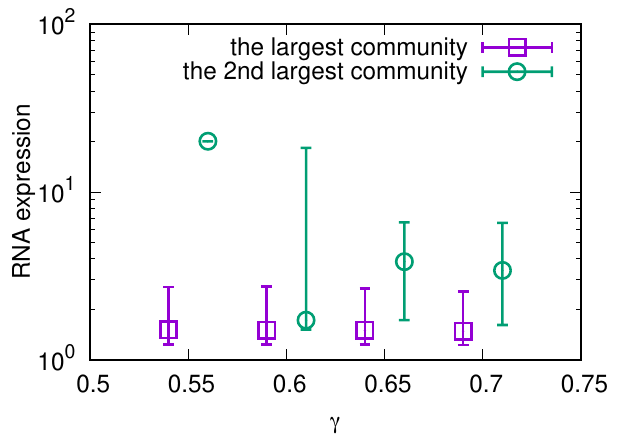} & \\
\end{tabular}
\caption{Differential RNA expression levels for different communities. We present the differential RNA expression for communities corresponding to various values of $\gamma$ for chromosomes $19$--$22$ and X, only for the largest and second largest communities. For visualization, we shift the data points slightly to the left for the largest community and right for the second largest community for each of the $\gamma$ values. The plots indicate the median values with the quartiles as the error bars.
}
\label{fig:RNAexp_for_communities_chr19_X}
\end{figure}

In Supplementary Figs.~\ref{fig:RNAexp_for_communities_chr1_6}--\ref{fig:RNAexp_for_communities_chr19_X}, for all of the chromosomes, we show the average coverage of RNA-seq reads within communities for different $\gamma$ values.

\subsection*{Chromatin state analysis}
In Supplementary Fig.~\ref{fig:comm_state}, we use the chromatin states to investigate what types of chromatin states are enriched in different communities at different $\gamma$ values.

\begin{figure}
\centering
\includegraphics[width=0.4\textwidth]{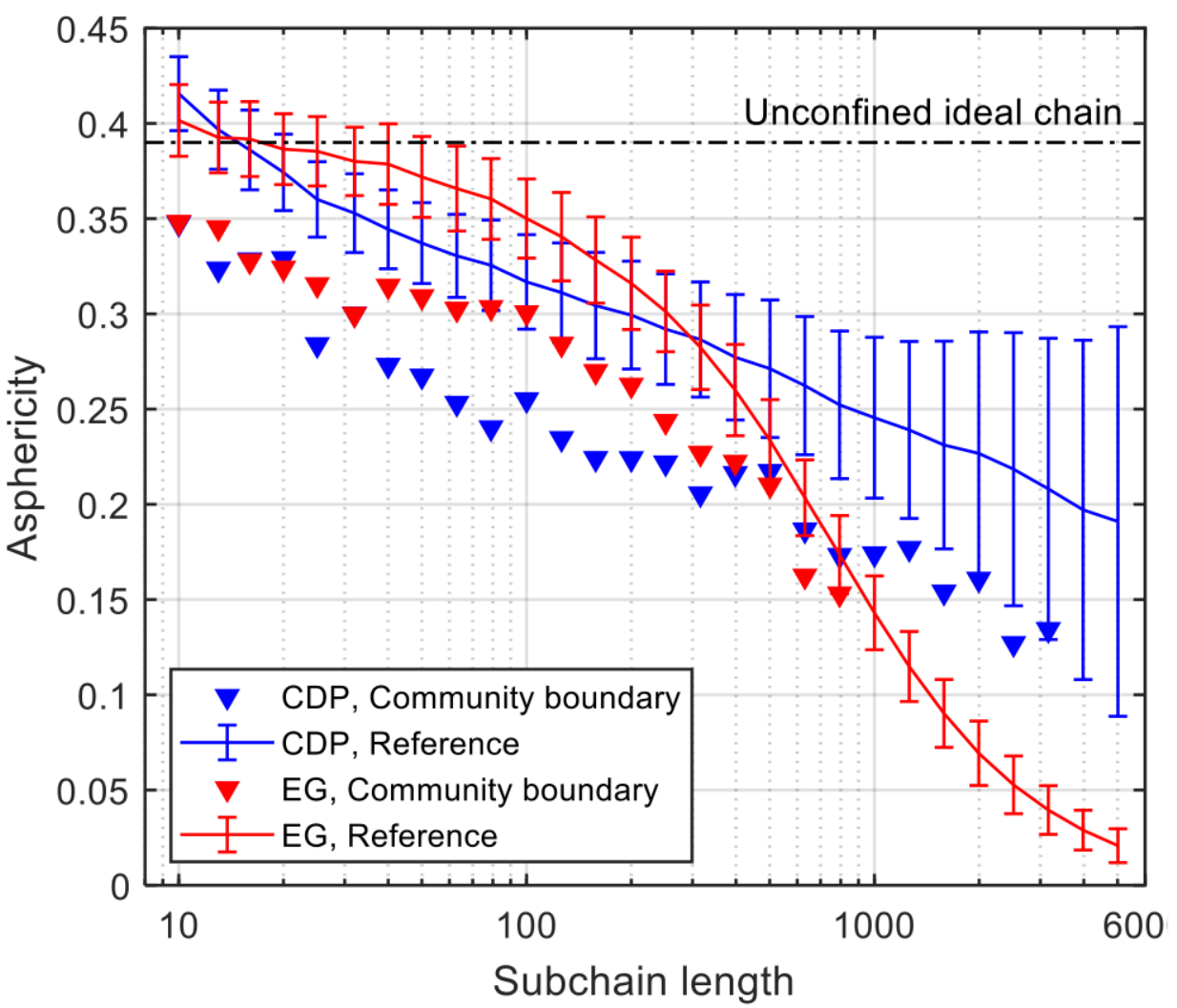}
\caption{The asphericity $\kappa^2$ as a function of subchain length for fractal (FG, blue) and equilibrium globules (EG, red) averaged over $200$ polymer realizations. The triangles denote the asphericity for the communities, and solid lines the polymer as a whole. To find the communities, we use $\gamma = 0.4$. The data were obtained from the simulation of $200$ sample globules for each polymer model. The error bars show the standard error of the mean. The dashed-dotted horizontal line corresponds to the value $\approx 0.39$ for unconfined ideal chain~\cite{rudnick1987shapes}.}
\label{fig:AsphericityEffectiveDistance}
\end{figure}

\begin{figure}
\centering
\begin{tabular}{cc}
chromosome 1 & chromosome 2 \\
\includegraphics[width=0.4\textwidth]{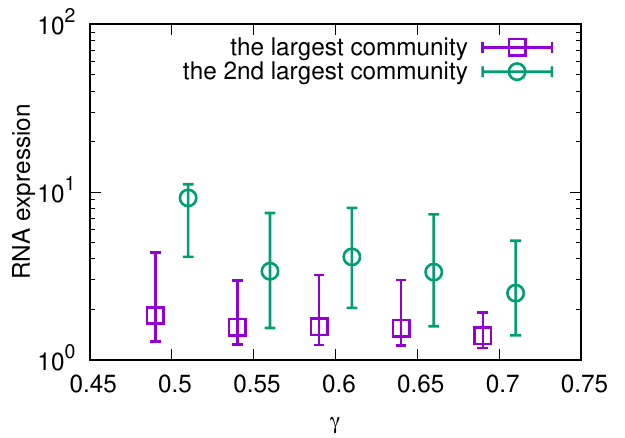} &
\includegraphics[width=0.4\textwidth]{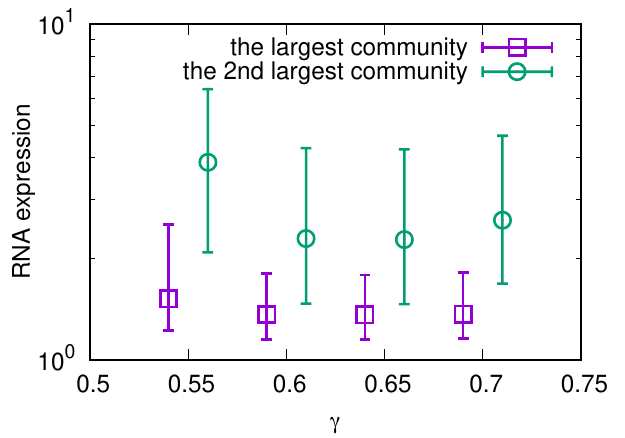} \\
chromosome 3 & chromosome 4 \\
\includegraphics[width=0.4\textwidth]{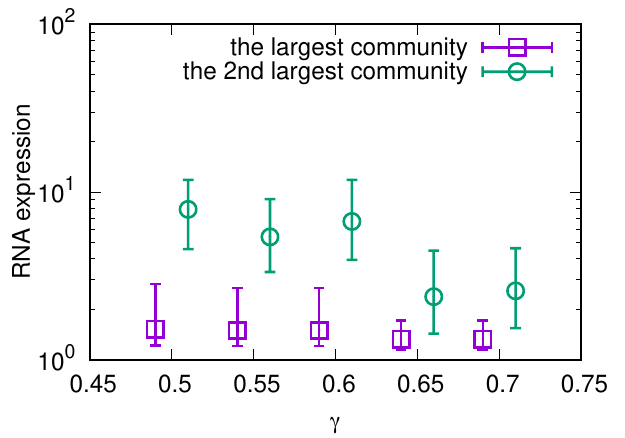} &
\includegraphics[width=0.4\textwidth]{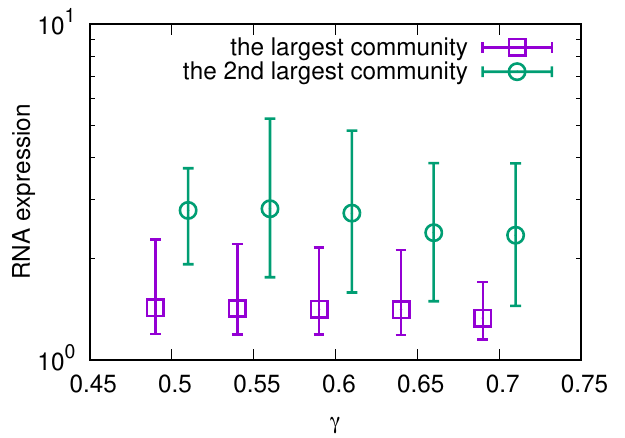} \\
chromosome 5 & chromosome 6 \\
\includegraphics[width=0.4\textwidth]{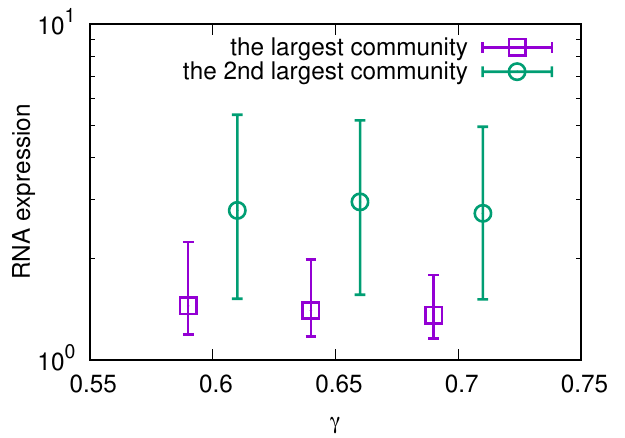} &
\includegraphics[width=0.4\textwidth]{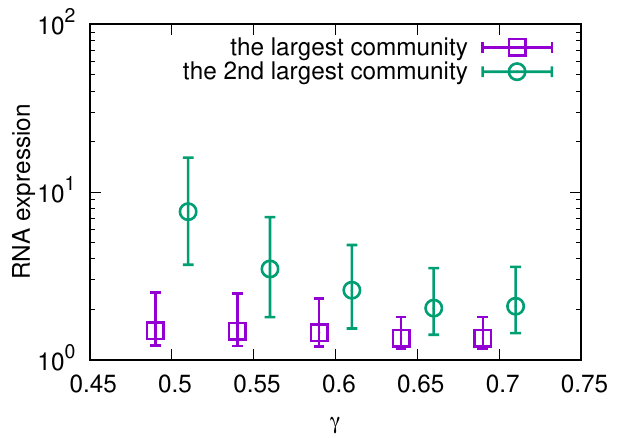} \\
\end{tabular}
\caption{Differential RNA expression levels for different communities. We present the differential RNA expression for communities corresponding to various values of $\gamma$ for chromosomes $1$--$6$, only for the largest and second largest communities. For visualization, we shift the data points slightly to the left for the largest community and right for the second largest community for each of the $\gamma$ values. The plots indicate the median values with the quartiles as the error bars.
}
\label{fig:RNAexp_for_communities_chr1_6}
\end{figure}

\begin{figure}[t]
\centering
\begin{tabular}{ll}
(a) & (b) \\
\includegraphics[width=0.4\textwidth]{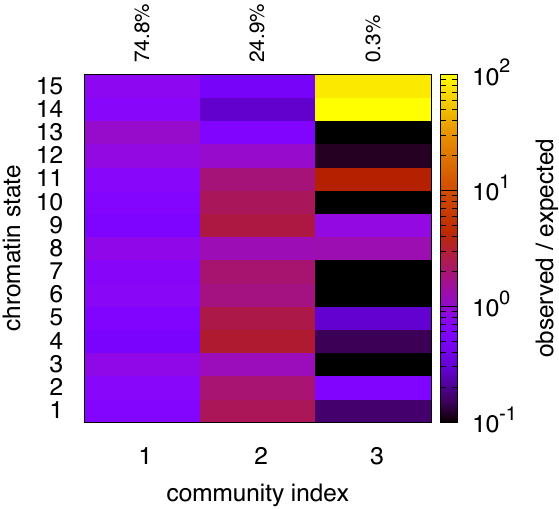} & 
\includegraphics[width=0.4\textwidth]{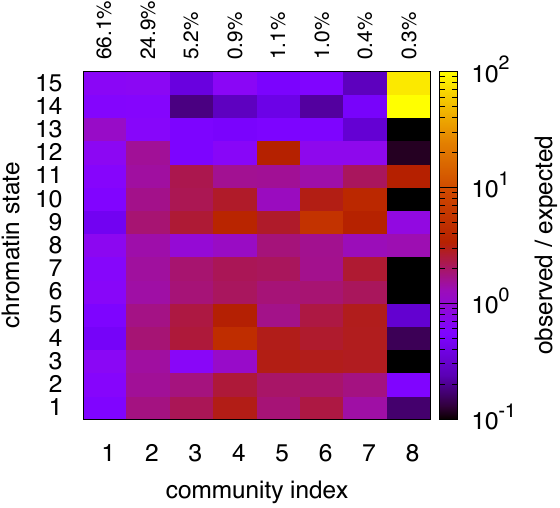} \\
(c) & (d) \\
\includegraphics[width=0.4\textwidth]{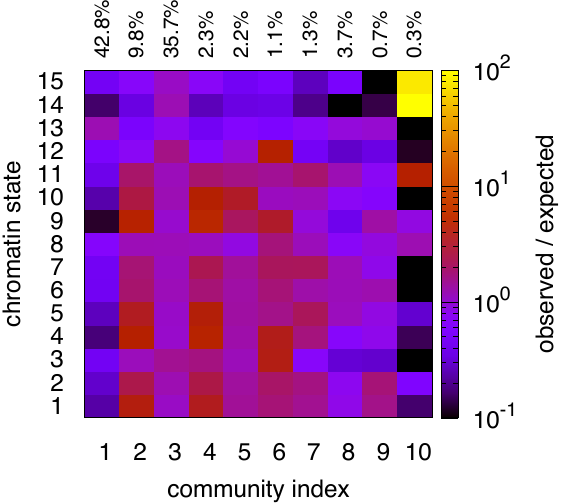} & 
\includegraphics[width=0.4\textwidth]{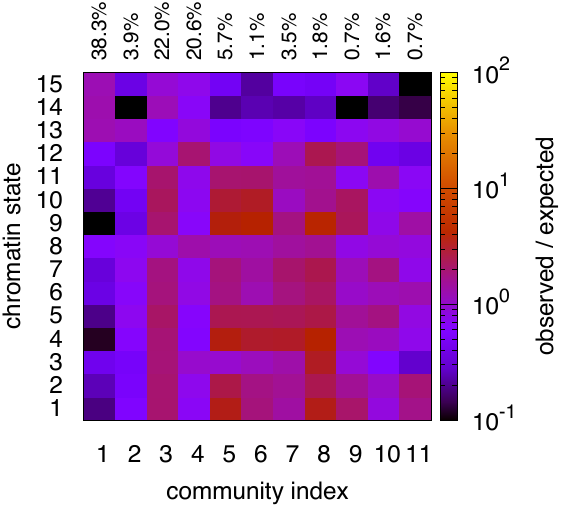} \\
\end{tabular}
\caption{The composition of chromatin segments for each community found for chromosome 1, in terms of chromatin states~\cite{Ernst2010,Ernst2011}. The color bar represents the fold difference between the total length of overlapping chromatin segments (in the unit of sequence) between the given community and the given chromatin state, and the expected overlap by assuming the random pairing between the communities and chromatin states. The percentage above each community index indicates the relative fraction of the community in the sequence. For all of the cases, $\chi^2 > 5 \times 10^7$ and the result is statistically significant with $p$-value $< 10^{-5}$. The values of the resolution parameter are (a) $\gamma = 0.6$, (b) $\gamma = 0.65$, (c) $\gamma = 0.7$, and (d) $\gamma = 0.75$. The chromatin states are 1: active promoter, 2: weak promoter, 3: poised promoter, 4: strong enhancer, 5: strong enhancer, 6: weak enhancer, 7: weak enhancer, 8: insulator, 9: transcriptional transition, 10: transcriptional elongation, 11: weak transcribed, 12: polycomb-repressed, 13: heterchromatin; low signal, 14: repetitive/copy number variation, and 15: repetitive/copy number variation.}
\label{fig:comm_state}
\end{figure}

\end{document}